\documentclass[a4paper,fleqn,usenatbib]{mnras}
\usepackage{multirow}
\usepackage{amsmath}
\usepackage{amssymb}
\usepackage{graphicx}   
\usepackage{psfrag}
\usepackage{epstopdf}
\makeatletter
\def\fps@figure{htbp} 
\makeatother

\def \apj  {ApJ}
\def \apjs  {ApJS}

\def \mnras {MNRAS} 
 
\def \apjl {ApJL} 
\def \L {\tilde{L} } 
\newcommand{\msun}{\,{\rm M_{\odot}}}
\newcommand{\cm}{\,{\rm cm}}

\title[Jet propagation in dense media]{Numerically calibrated model for propagation of a relativistic unmagnetized jet in dense media}
\author[Harrison, Gottlieb \& Nakar]{Richard Harrison, Ore Gottlieb  and Ehud Nakar\thanks{udini@wise.tau.ac.il}
\\
The Raymond and Beverly Sackler School of Physics and Astronomy, Tel Aviv University, Tel Aviv 69978, Israel\\
}

\begin{document}
\maketitle	
	
\begin{abstract}
Relativistic jets reside in high-energy astrophysical systems of all scales. Their interaction with the surrounding
media is critical as it determines the jet evolution, observable signature, and feedback on the environment. During its motion the interaction of the jet with the ambient media inflates a highly pressurized cocoon, which under certain conditions collimates the jet and strongly affects its propagation. Recently, \cite{bromberg2011} derived a general simplified (semi)analytic solution for the evolution of the jet and the cocoon in case of an unmagnetized jet that propagates in a medium with a range of density profiles. In this work we use a large suite of 2D and 3D relativistic hydrodynamic simulations in order to test the validity and accuracy of this model. We discuss the similarities and differences between the analytic model and numerical simulations and also, to some extent, between 2D and 3D simulations. Our main finding is that although the analytic model is highly simplified, it properly predicts the evolution of the main ingredients of the jet-cocoon system, including its temporal evolution and the transition between various regimes (e.g., collimated to uncollimated). The analytic solution predicts a jet head velocity that is faster by a factor of about 3 compared to the simulations, as long as the head velocity is Newtonian. We use the results of the simulations to calibrate the analytic model which significantly increases its accuracy. We provide an applet that calculates semi-analytically the propagation of a jet in an arbitrary density profile defined by the user at \url{http://www.astro.tau.ac.il/~ore/propagation.html}.
\end{abstract}


\section{Introduction}
Relativistic jets are present in many astrophysical systems where there is accretion on a compact object (e.g., gamma-ray bursts, tidal disruption events, active galactic nuclei  and $\mu$-quasars). Often these jets are propagating within a medium which is dense enough to affect their dynamics (e.g., a stellar envelope in the context of long gamma-ray bursts). The properties of these jets vary by orders of magnitude in size, duration and energy content, yet the physical principles that dictate their evolution may be similar. In this paper we study these principles numerically for the propagation of a relativistic hydrodynamic (unmagnetized) jet in a cold, static and spherically symmetric medium.

The propagation of relativistic hydrodynamic jets was explored by many authors both analytically \citep[e.g.,][]{blandfordrees74,begelman1989,meszaros2001,Matzner03,lazzati2005} and numerically  \citep[e.g.,][]{marti1995,marti1997,aloy2000,macfadyen2001,reynolds2001,zhang2004,mizuta2006,morsony2007,wang2008,lazzati2009,mizuta2009, morsony2010,nagakura2011,mizuta2013,lopez-camara2013,ito2015,lopez-camara2016}. The picture that arises from these studies is that the jet propagation forms a double bow-shock head structure. Because of a high pressure gradient across the jet head, material that enters the head is forced out perpendicular to the propagation of the head, creating a hot cocoon that surrounds the jet. This hot cocoon applies pressure on the jet and possibly leads to its collimation. 

Recently, \cite{bromberg2011} (hereafter B11) derived a general simplified solution for the propagation of a relativistic unmagnetized jet in an external medium with a range of density profiles. It provided a self-consistent analytic solution to the equations of the jet-cocoon system and its evolution with time. While being useful in understanding and analyzing the propagation of such jets, the analytic solution has also its limitations. First, it makes assumptions that can be only partially justified analytically. Second, being simplified the solution provides only order of magnitude estimates.  The first purpose of this paper is to use numerical simulations in order to test the analytic solution of B11. Namely, to check the validity of their assumptions and to verify that the evolution in the different regimes discussed by B11 indeed follows their solution. The second goal of this paper is to calibrate the values of the order-of-magnitude coefficients in the B11 solution in order to obtain a more accurate, yet simple, analytic description of the jet propagation.

For that we employ a large set of numerical simulations in 2D and 3D with a variety of jet and external medium properties. Below, for completeness, we first present an outline of the model presented by B11 (\S\ref{sec:mod}). Then we describe the simulations' configurations (\S\ref{sec:num}), followed by an analysis of the simulations' results  (\S\ref{sec:res}).  In \S\ref{sec:calibration} we use these results to derive a calibrated analytic model. In \S\ref{sec:star} we study the specific setting of a jet propagation in a stellar envelope, comparing the calibrated analytic model to 3D simulations.
In the appendix we provide the analytic solutions of the various regimes for a medium with a power-law density profile. An applet that calculates the propagation of a jet in an arbitrary density profile defined by the user is provided at \url{http://www.astro.tau.ac.il/~ore/propagation.html}.

 
\section{Analytic Model}
\label{sec:mod} 

\begin{figure}
\begin{center}
\includegraphics[width=1\columnwidth]{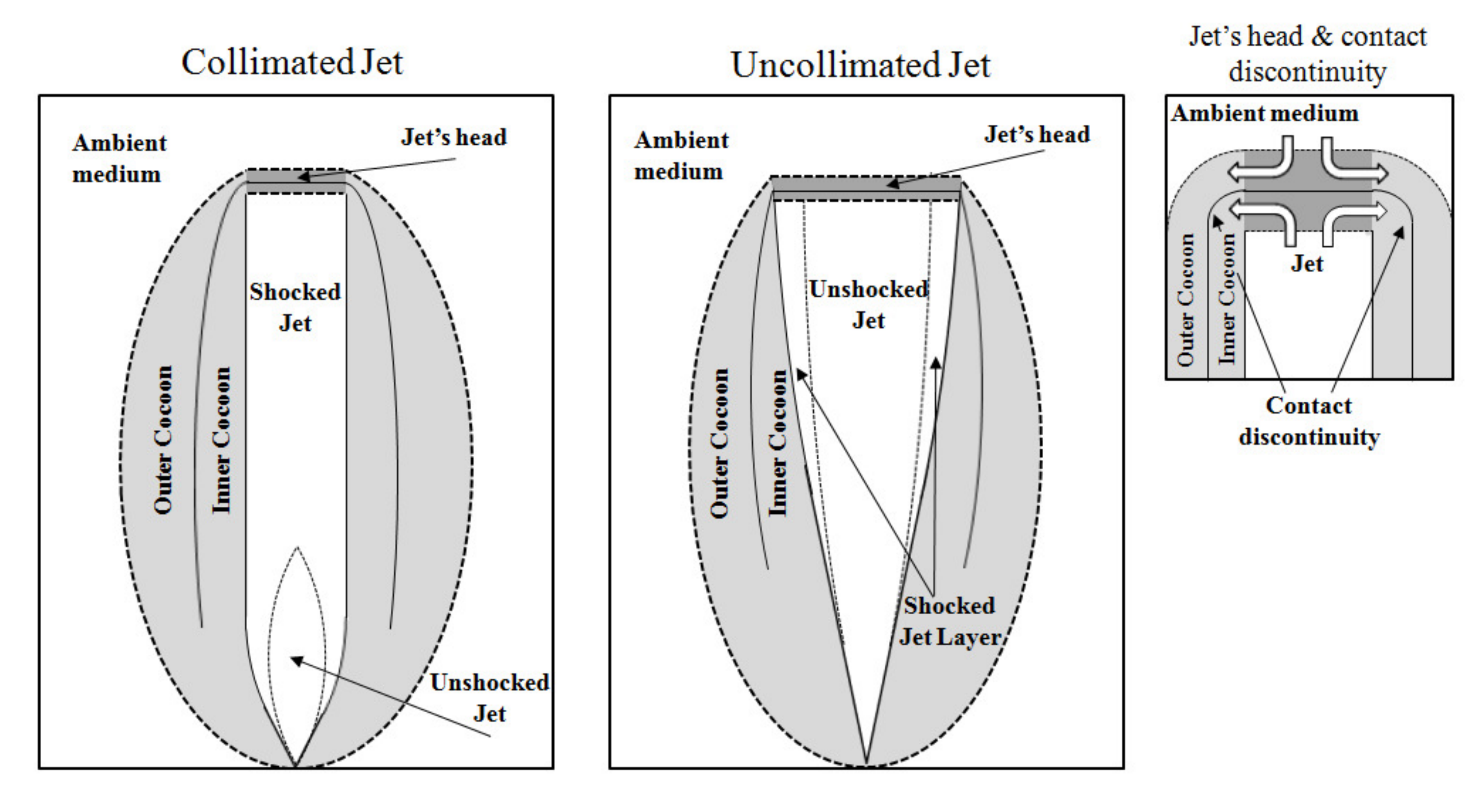}
\caption{Image from B11 showing the jet geometry for a collimated ({\it left}) and an uncollimated ({\it middle}) jet. Both regimes show the same basic properties of the jet with its head propagating into the ambient medium. The jet head (enlarged in the right panel) feeds material into the cocoon which consists of an inner light shocked jet cocoon and an outer heavy shocked ambient material cocoon. The primary difference between the two cases is that in the collimated case the collimation shock closes upon the axis, where the jet becomes cylindrical, while in the uncollimated case the collimation shock does not converge and the jet is approximately conical up to the head.
}
\label{fig:schem}
\end{center}
\end{figure}

We  give here a brief  overview of the  jet model presented  in B11, we direct  the reader to  this work for  a detailed
description. The model considers  an axis-symmetric configuration were an unmagnetized cold  relativistic jet with a
{\it one sided}\footnote{The total luminosity of a bi-polar jet is $2L_j$ and the isotropic equivalent luminosity of the jet is $4L_j/\theta_0^2$. } luminosity $L_j$ and an  opening angle $\theta_0 \ll 1$ rad is injected into a static ambient medium with a density profile
$\rho_a(z)$ and a negligible pressure. It  is applicable to a density profile that is  not dropping too fast, specifically that $\rho_a z^3$ grows monotonically with $z$. { For example if  $\rho_a(z) \propto z^{-\alpha}$ then the solution is applicable only for $\alpha<3$.} The jet propagates in the  $z-$direction in cylindrical coordinates and
is approximated to  be axisymmetric along that axis.  At the leading edge of  the jet a forward shock  heats the ambient
medium and a  reverse shock heats the jet.  The two shocks are  separated by a contact discontinuity.  This double shock
system is called the jet head.  The pressure of the ambient medium is low compared to  that of the shocked head material
leading the  head material to flow  sideways, i.e., away  from the axis in  the $r-$direction. This forms  a pressurized
cocoon that surrounds the  jet. The hot cocoon, in turn, applies  pressure both on the cold jet and  on the cold unshocked
ambient medium.  This leads to a  forward shock that propagates  in the $r$ direction  into the ambient medium  and to a
collimation shock that propagates into the cold jet. The latter  may lead, depending on the system parameters, to a full
collimation of the  jet (see schematic Figure \ref{fig:schem}).  The analytic solution solves for the  head velocity and
location, cocoon  pressure and  size, the  jet collimation  shock height  and  the shocked  jet Lorentz  factor and
cross-section.

B11 show that almost all the jet and cocoon properties are determined (up to coefficients that depend on the exact density profile) by only two dimensionless parameters, the jet opening angle, $\theta_0$, and  the ratio of jet energy density and rest-mass energy density of the ambient medium at the head
\begin{equation}\label{eq:Ltild}
	\tilde{L}\equiv\frac{\rho_jh_j\Gamma_j^2}{\rho_a}\approx\frac{L_j}{\Sigma_j\rho_ac^3},
\end{equation}
where $\rho_j$, $h_j$, $\Gamma_j$ and $\Sigma_j$ are the jet mass density, specific enthalpy, Lorentz factor and head cross-section, respectively.  $\tilde{L}$ depends on the jet cross-section, which depends on the level of the jet  collimation, which in turn depends on the cocoon pressure and the head location, both depending on $\tilde{L}$. The solution of B11 is a closed set of equations that simultaneously solves for all these parameters self-consistently.

The jet evolution depends first on whether it is collimated or not. The criterion for collimation is: 
\begin{equation}\label{eq:col}
\tilde{L} \theta_0^{4/3}\left(\frac{16\Omega}{3}\right)^{-2/3} < 1,
\end{equation}
where $\Omega$ is a coefficient that depends on several integrals over the evolution of the jet and cocoon and can vary by a factor of a few (see appendix \ref{app:intParam} for a full description of $\Omega$). The expression for $\L$ depends on the regime, where for a collimated jet it is
\begin{equation}
\tilde{L}= \left(\frac{Lj}{\theta_0^4 t^2\rho_a(z_h) c^5}\right)^{2/5}\left(\frac{16\Omega}{3\pi}\right)^{2/5}~~~~{\rm (collimated)} ,
\label{eq:Ltcol}
\end{equation}
and for an uncollimated jet it is 
\begin{equation}
\tilde{L}=\frac{L_j}{\pi \theta_0^2 t^2\rho_a(z_h) c^5}~~~~{\rm (uncollimated)}.
\label{eq:Ltuncol}
\end{equation}
where $\rho_a(z_h)$ is the ambient density at the head location.  
$\L$ determines the head propagation velocity \citep{Matzner03}
\begin{equation}
\beta_h~=~\frac{1}{1+\tilde{L}^{-1/2}}~,
\label{eq:hvel}
\end{equation}
and together with other parameters it gives the cocoon pressure 
\begin{equation}
P_c=\left(\frac{L_j\rho_a}{3\pi c}\right)^{1/2}\tilde{L}^{-1/4}t^{-1}\Omega^{1/2},
\end{equation}
and the cocoon sideways expansion velocity
\begin{equation}\label{eq:beta_c}
\beta_c=\sqrt{\frac{P_c}{\bar{\rho}_a(z)c^2}}
\end{equation}
where $\bar{\rho}_a$ is the average external medium density in the  volume occupied by the cocoon. The collimation shock height (in the collimated regime) is
\begin{equation}
  \hat{z}=2r_j/\theta_0	
\end{equation}
where $r_j=(\Sigma_j/\pi)^{1/2}$ is the jet radius after collimation. Note that in order to obtain this closed set of equations one also needs an equation for  the jet cross-section:
\begin{eqnarray}
\Sigma_j&=&\frac{L_j\theta_0^2}{4cP_c} ~~ (collimated)\nonumber \\
&&  \\
\Sigma_j&=&\pi z_h^2\theta_0^2 ~~ (uncollimated). \nonumber
\end{eqnarray}

In general this closed set of propagation equations must be integrated numerically.  However in cases that the head is either fully { Newtonian ($ \L \ll 1 $) or relativistic ($ \L \gg 1 $),} its velocity and Lorentz factor can be approximated as:
\begin{equation}	
	\begin{array}{lccr}
		\beta_h & \approx & \L^{1/2} &~~~\rm{for}~\L \ll 1 \\
		&&&\\
		\Gamma_h & \approx & \frac{1}{\sqrt{2}}\L^{1/4} & ~~~\rm{for}~\L \gg 1 
	\end{array} ,
\label{eq:hvel_approx}  
\end{equation}
were $\Gamma_h=(1-\beta_h^2)^{-1/2}$ is the head Lorentz factor. If in addition the external
density profile can be approximated by a power-law, $\rho_a \propto z^{-\alpha}$, then the integration parameter is
\begin{equation}\label{eq:Omega}
	\Omega= ֿ\left\{ 
	\begin{array}{lr}
	  \frac{81}{(5-\alpha)^3(3-\alpha)} & ~~~\L \ll 1 \\
	   &\\
	  \frac{3(3+\alpha)^2}{5(3-\alpha)(7-\alpha)}   &  ~~~\L \gg 1 
	\end{array}
	\right.
\end{equation}
and the entire system can be solved analytically. In appendix \ref{app:analyticModel} we provide an analytic solution of the equations in the different regimes (including the calibration based on numerical simulation results).

In deriving the above equations B11 make several assumptions and approximations. The main assumption is that the cocoon pressure can be well approximated as being uniform. This is expected in case that the entire cocoon is causally connected, which requires that the shocked jet material in the cocoon (inner cocoon in Figure 1) is not well mixed with the shocked external medium. Without strong mixing the shocked jet material that fills the cocoon is hot enough so its sound speed is faster than, or at least comparable to, the head velocity. The cocoon mixing is also important to determine the fate of the cocoon material after the jet breaks out of its surrounding media (e.g., stellar envelope) and the material spills out \citep[e.g.,][]{ramirez-ruiz2002,lazzati2010,nakar2017,gottlieb2017a}.

B11 make several additional simplifying approximations such as that all the energy injected into the head in the collimated regime spills into the cocoon, a uniform ram pressure across the head, a cylindrical shape for the shocked jet and the cocoon, etc. All these are important mostly in the collimated case, where the cocoon properties drives the jet structure. This is in contrast to the uncollimated regime where the feedback of the cocoon on the jet structure and evolution is minor (most of the jet material is moving ballistically between the launching site and the head and it is well approximated by a freely expanding jet). Thus, in this work, which aims at testing and calibrating the B11 model, we focus on the collimated regime and the transition to the un-collimated regime.

\section{Simulation settings} 
\label{sec:num}
In order to test numerically the assumptions of the analytic model and to calibrate its coefficients we take a clean setting with a  power-law density profile\footnote{Since the jet and the cocoon are  narrowly collimated around the symmetry axis, $z$, the density profile that we consider, $\rho_a \propto z^{-\alpha}$, is practically similar to a spherically symmetric density profile.} $\rho_a \propto z^{-\alpha}$ and a jet with constant properties (i.e., $L_j$ and $\theta_0$).
Under these conditions the case of $\alpha=2$ is unique in the sense that $\L$ is constant in time. Thus, all the velocities in the system (e.g. $\beta_h$, $\beta_c$) are constant and all the quantities with dimensions of length grow linearly with time. For $\alpha<2$, $\tilde{L}$ decreases with time and the jet decelerates, while for $\alpha>2$, $\tilde{L}$ increases with time and the jet accelerates.
The $\alpha=2$ case is convenient for our tests since the transition between the various regimes depends on $\L$ and $\theta_0$ and having both constant makes it simpler to test. We also expect that the numerically derived calibration coefficients may depend on $\L$, which again is most simple to derive in $\alpha=2$ settings. We therefore run most of the simulations with $\alpha=2$ and verify the results for other values of $\alpha$. We also carry out one simulation where $L_j$ varies in time as a power-law, to test the model for a varying luminosity and one 3D simulation of a jet within a density profile of a stellar envelope to test the semi-analytic model on a non power-law density distribution.

We use PLUTO code version 4.0, which is a freely distributed numerical code \citep{mignone2007,mignone2012}. The code is designed to integrate a general system of conservation laws, and we use its special relativistic hydrodynamics module. Our 2D simulations use a cylindrical coordinate system with a static grid $z-r$ while in 3D we use Cartesian coordinates. In both the jet is launched along the $z-\mathrm{axis}$. Throughout this work we apply an ideal equation of state with a constant polytropic index of $4/3$, as appropriate for a radiation dominated gas. {We verify that there are no significant differences in case that the gas pressure is not dominated by radiation and the equation of state vary between the relativistic and Newtonian regimes (see \ref{app:eos}).} For the jet injection in these coordinate systems we utilized the technic used by \cite{mizuta2013}. Instead of injecting a conical cold jet we inject a relativistically hot cylindrical jet with a bulk Lorentz factor $\Gamma_{j,0}$ through a nozzle with a radius $r_{noz}$ parallel to the $z-\mathrm{axis}$ at some height $z_{base}$. Being relativistically hot the jet accelerates and expands sideways quickly to form a cold conical jet with a roughly uniform luminosity per unit of solid angle over an opening angle $\theta_0=1/(f\Gamma_{j,0})$, where $f$ is a factor of order unity. By examining the jet structure we measure\footnote{ $f$ relates the injection Lorentz factor to the asymptotic jet opening angle once it expands and accelerate after being injected from a nozzle and before it starts being affected by the interaction with the surrounding media. Note that this is a different quantity than the one measured by \cite{mizuta2013}, which discuss the asymptotic opening angle after breakout following the jet crossing of  a stellar envelope}. $f\approx 1.4$. For both 2D and 3D simulations we chose a nozzle size of $r_{noz}=10^{8}\mathrm{cm}$ allowing sufficient mesh coverage over the nozzle and $z_{base}=10^9$ cm. This is significantly smaller than the jet radius and collimation height we expect in our simulations and thus should not affect the jet dynamics. We verify that this is the case (see appendix \ref{app:nozzle}). We perform detailed convergence tests of our simulations. We achieve a convergence of the jet propagation speed and of the jet structure. We discuss these tests in appendix \ref{app:convergence}.

\subsection{2D simulations}
In most simulations we use a constant canonical luminosity\footnote{The jet dynamics depends on the ratio $L_j/\rho_a$ as we vary $\rho_a$ there is no need to vary $L_j$ as well.} ($L_j=5\times10^{50}\mathrm{erg/s}$) and the same specific enthalpy for the injected jet, $h_{j,0}=1+4P_{j,0}/\rho_{j,0} c^2=110$. We carry out also a simulation where $L_j$ varies as a power-law in time and two simulations with $h_{j,0}=20,\, 200$ to see its effect. The density normalization at $z_{base}$, the density power-law $\alpha$, and the injection Lorentz factor $\Gamma_{j,0}$ (which translates to the value of $\theta_0$) are varied between the different simulations. The properties of all the 2D simulations we ran are given in table \ref{tab:sim}.

As a canonical grid we take 4000 grid points in the $z-$direction and 400 in the $r-$direction. We split the grid space into two regions, a uniform mesh distribution near the jet axis and at small $z$, and a logarithmic mesh spacing at larger distances. The uniform meshes cover $0<r<2\times10^9\mathrm{cm}$ ($\Delta r=10^{7}\mathrm{cm}$) and $1\times10^9<z<4\times10^{10}\mathrm{cm}$ ($\Delta z=2\times10^{7}\mathrm{cm}$)\footnote{{We verify that using rectangular cells with an aspect ratio 2:1 ($\Delta z=2 \Delta x$) does not affect our result by running simulation 12 with $\Delta z=\Delta x$ and finding no significant diferences.}}. The logarithmically spaced meshes distributed in the range $2\times10^9<r<1\times10^{11}\mathrm{cm}$ and $4\times10^{10}<z<4\times10^{11}\mathrm{cm}$ (the logarithmic spacing coefficient is $2\times10^{-4}$). {The resolution in our canonical grid it therefore comparable to previous 2D studies \citep{lazzati2009,nagakura2011,lazzati2012,mizuta2013}.}

\subsection{3D simulations}
In all 3D simulations we use a constant $L_j=5\times10^{50}\mathrm{erg/s}$ and $h_{j,0}=110$. 
We make use of a Cartesian coordinate system with the jet propagating in the $z-$direction.
For simulations 3Da-3De we divide the grid to five patches along the $ x $ and $ y $ axes and two patches on the $ z $-axis. On the axes perpendicular to the jet, for the innermost region that covers the jet spine between $ -2 \times 10^8 $cm and $ 2 \times 10^8 $cm we implement a high resolution cells size of $ 1.66\times 10^7 $cm. Outside to this region the patches are symmetric and distributed logarithmically with respect to the the $ z $ axis, the inner patch has 32 cells up to $ |10^9| $cm, while the outer patch contains 70 cells up to $ |2\times 10^{10}| $cm. On the $ z $-axis the first patch is from $ 10^9 $cm to $ 2\times 10^{10} $cm with 360 cells that are distributed uniformly. The second patch is up to $ 10^{11} $cm with 600 cells distributed logarithmically. The total number of cells is therefore $ 228 \times 228 \times 960 $. {This resolution is slightly higher than most previous 3D studies \citep{wang2008,lopez-camara2013,ito2015}.}
We also implement an higher resolution simulation, 3Df. On the $ x $ and $ y $ axes the innermost patch between $ -10^9 $cm and $ 10^9 $cm has a uniform distribution of 150 cells and the outer patch, symmetric with respect to the $ z $-axis, stretched up to $ |5\times 10^{10}| $cm with 200 cells distributed logarithmically. On the $ z $-axis the patches are the same as for simulations 3Da-3De, but with 570 and 1200 cells on the first and second patches, respectively. The total number of cells is $ 550 \times 550 \times 1770 $. For both grids we carried out convergence tests as described in appendix \ref{app:convergence}. 


\begin{table}
\caption{2D Simulations}
\centering
\begin{tabular}{ |c|c|c|c|c|c|}
\hline
\multicolumn{6}{ |c| }{$L_j=5\times10^{50}\mathrm{erg/s}$ (one sided jet luminosity)} \\
\hline
Model & $\alpha$ & $\rho_a(z=10^9{\rm cm})$ & $\theta_0$ & $\tilde{L}_s$ & $h_{j,0}$ \\
&  & $[\mathrm{g/cm^3}]$ & [rad] & &\\ \hline
1 & \multirow{4}{*}{0} & $10^5$ & 0.14 & 0.0008-0.004 & 110 \\ 
2 & & $10^4$ & 0.14 & 0.0005-0.004 & 110 \\
3 & & $10^3$ & 0.14 & 0.005-0.01 & 110 \\
4 & & $10^2$ & 0.14 & 0.02-0.3 & 110 \\ \hline
5 & \multirow{4}{*}{1} & $10^5$ & 0.14 & 0.006-0.01 & 110 \\ 
6 & & $10^4$ & 0.14 & 0.02-0.04 & 110 \\
7 & & $10^3$ & 0.14 & 0.06-0.1 & 110 \\
8 & & $10^2$ & 0.14 & 0.3-0.7 & 110 \\ \hline
9 & \multirow{10}{*}{2} & $10^7$ & 0.14 & 0.001 & 110\\ 
10 & & $10^6$ & 0.14 & 0.007 & 110 \\
11 & & $10^5$ & 0.14 & 0.04 & 20 \\
12 & & $10^5$ & 0.14 & 0.04 & 110 \\
13 & & $10^5$ & 0.14 & 0.04 & 200 \\
14 & & $10^4$ & 0.14 & 0.35 & 110 \\
15 & & $10^3$ & 0.14 & 1.7 & 110 \\
16 & & $10^2$ & 0.14 & 25 & 110 \\
17 & & $10^1$ & 0.14 & 47 & 110\\
18 & & $10^0$ & 0.14 & 447 & 110\\ 
19 & & $10^4$ & 0.18 & 0.16 & 110\\
20 & & $10^5$ & 0.07 & 1.5 & 110\\
21 & & $10^3$ & 0.07 & 22 & 110\\
22 & & $10^5$ & 0.04 & 4.5 & 110\\
23 & & $10^3$ & 0.04 & 75 & 110\\
24 & & $10^1$ & 0.04 & 510 & 110\\ \hline
25 & \multirow{4}{*}{2.5} & $10^5$ & 0.14 & 0.3-0.37 & 110 \\ 
26 & & $10^4$ & 0.14 & 1.6-2.5 & 110 \\
27 & & $10^3$ & 0.14 & 6.4-10 & 110 \\
28 & & $10^2$ & 0.14 & 10-72 & 110\\ \hline
\multicolumn{6}{ |c| }{$L_j=5\times10^{50} \left(\frac{t}{1 {\rm ~s}}\right)^{0.5} \mathrm{erg/s}$} \\ \hline
29 & 1.5 & $10^5$ & 0.14 & 0.04 & 110 \\ \hline
\end{tabular}
\label{tab:sim}
\begin{flushleft}
Simulation properties: Model number; ambient medium density: power-law index ($\alpha$) and normalization at $z=10^9$ cm; jet initial opening angle ($\theta_0$); the range of $\L$ probed in the simulation (see section \ref{sec:calibration} for the definition of $\L_s$); the initial specific enthalpy ($h_{j,0}$).
\end{flushleft}
\end{table}
\begin{table}
\caption{3D Simulations}
\setlength{\tabcolsep}{4pt}
\centering
\begin{tabular}{ |c|c|c|c|c|c|c|c|}
\hline
	\multicolumn{8}{ |c| }{$L_j=5\times10^{50}\mathrm{erg/s}~,~h_{j,0}=110$} \\
\hline
Model & $\alpha$ & $\rho_a(z=10^9{\rm cm})$ & $\theta_0$ & $\tilde{L}_s$ & $ \beta_h $ & $ \beta_h $ \\ 
&  & $[\mathrm{g/cm^3}]$ & [rad] & & (3D) & (2D) & \\ \hline
3Da & 2 & $10^4$ & 0.18 & 0.41 & 0.39 & 0.28 \\
3Db & 2 & $10^4$ & 0.14 & 0.92 & 0.49 & 0.4 \\
3Dc & 2 & $10^5$ & 0.14 & 0.048 & 0.18 & 0.16 \\
3Dd & 1 & $10^4$ & 0.14 & 0.016 & 0.11 & 0.12 \\
3De  & 2.5 & $10^5$ & 0.14 & 0.73 & 0.46 & 0.38 \\
3Df  & 2 & $6\times 10^4$ & 0.14 & 0.10 & 0.24 & -- \\ \hline

\end{tabular}
\label{tab:sim3d}
\end{table}
\section{Numerical results and comparison to the analytic model}
\label{sec:res}
Below we present the results obtained by the set of numerical simulations we carried out (tables 1 \& 2). We discuss different aspects of the results, all in comparison to the predictions of the analytic model of B11. 


\begin{figure}
\begin{minipage}[b]{\columnwidth}
\centering
\includegraphics[width=.45\textwidth]{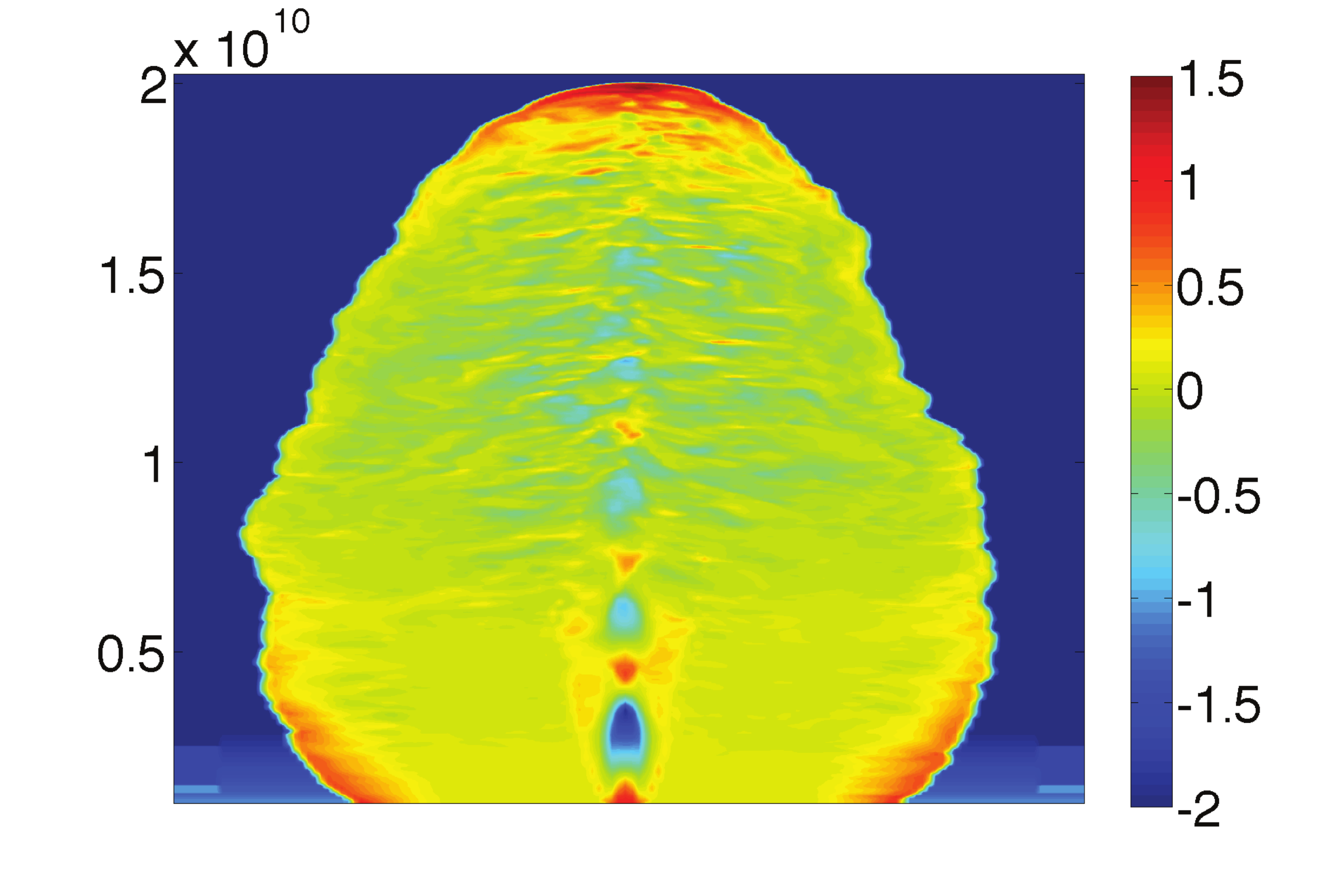}\quad
\includegraphics[width=.45\textwidth]{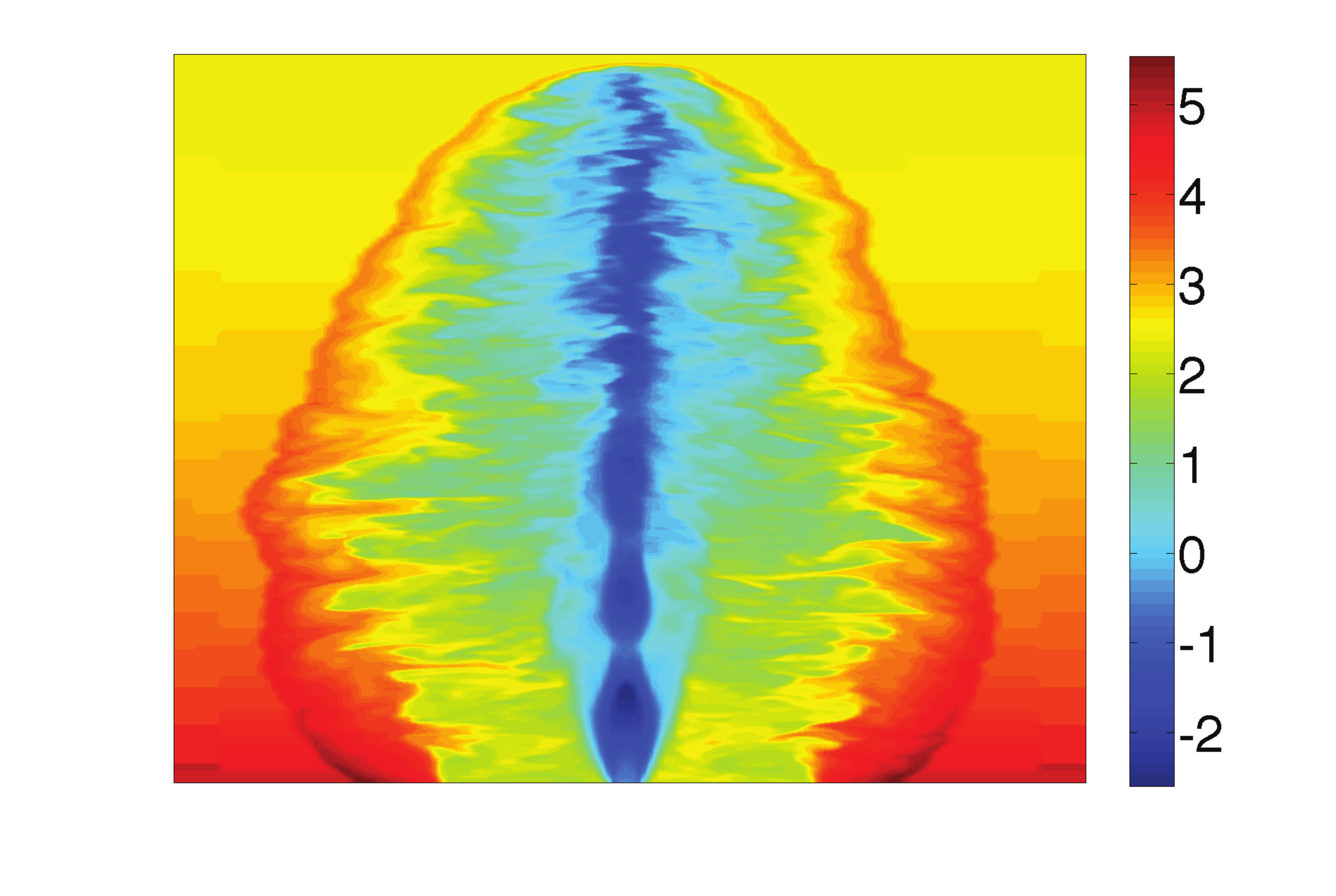}
\end{minipage} \hfill
\begin{minipage}[b]{\columnwidth}
\centering
\includegraphics[width=.45\textwidth]{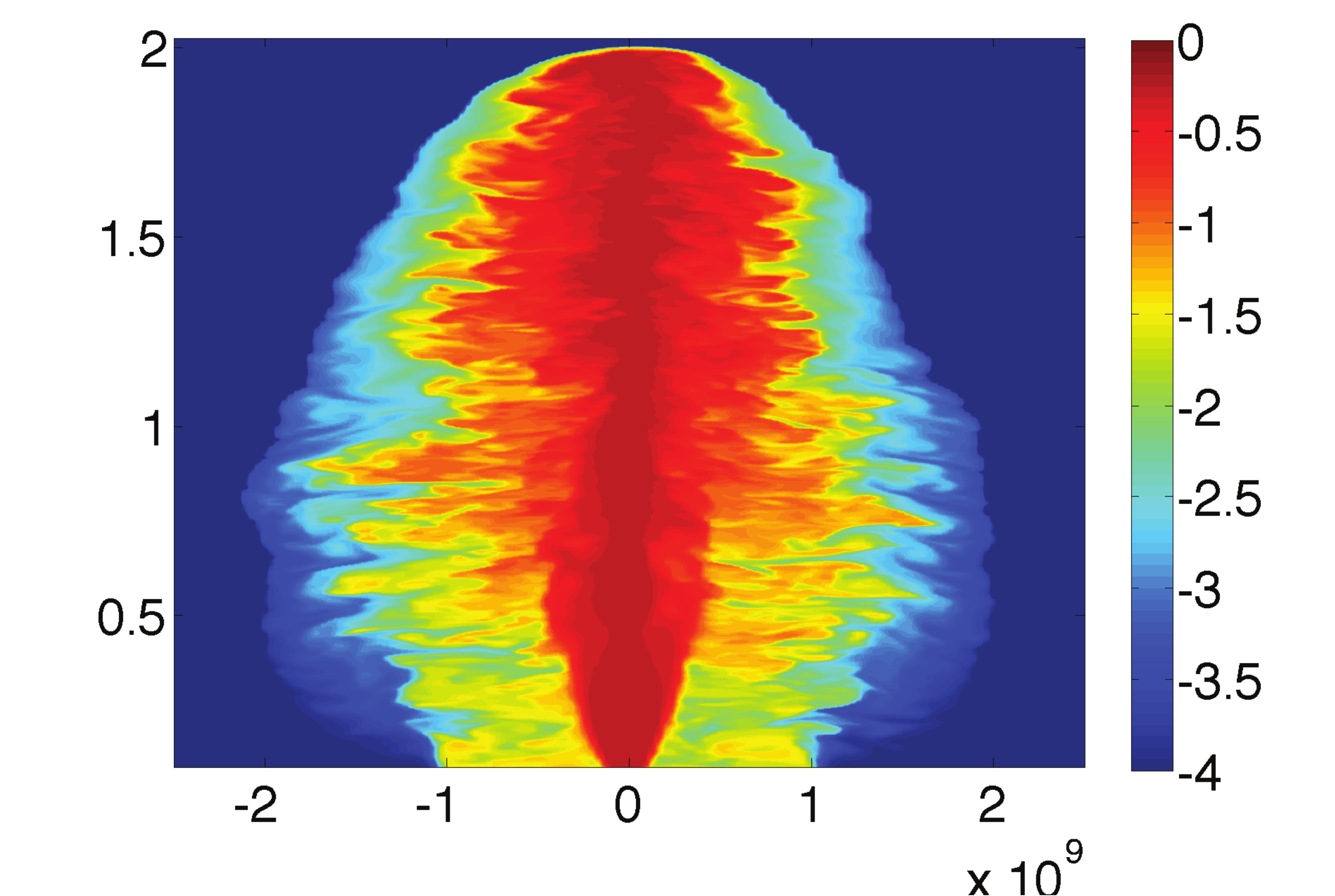}\quad
\includegraphics[width=.45\textwidth]{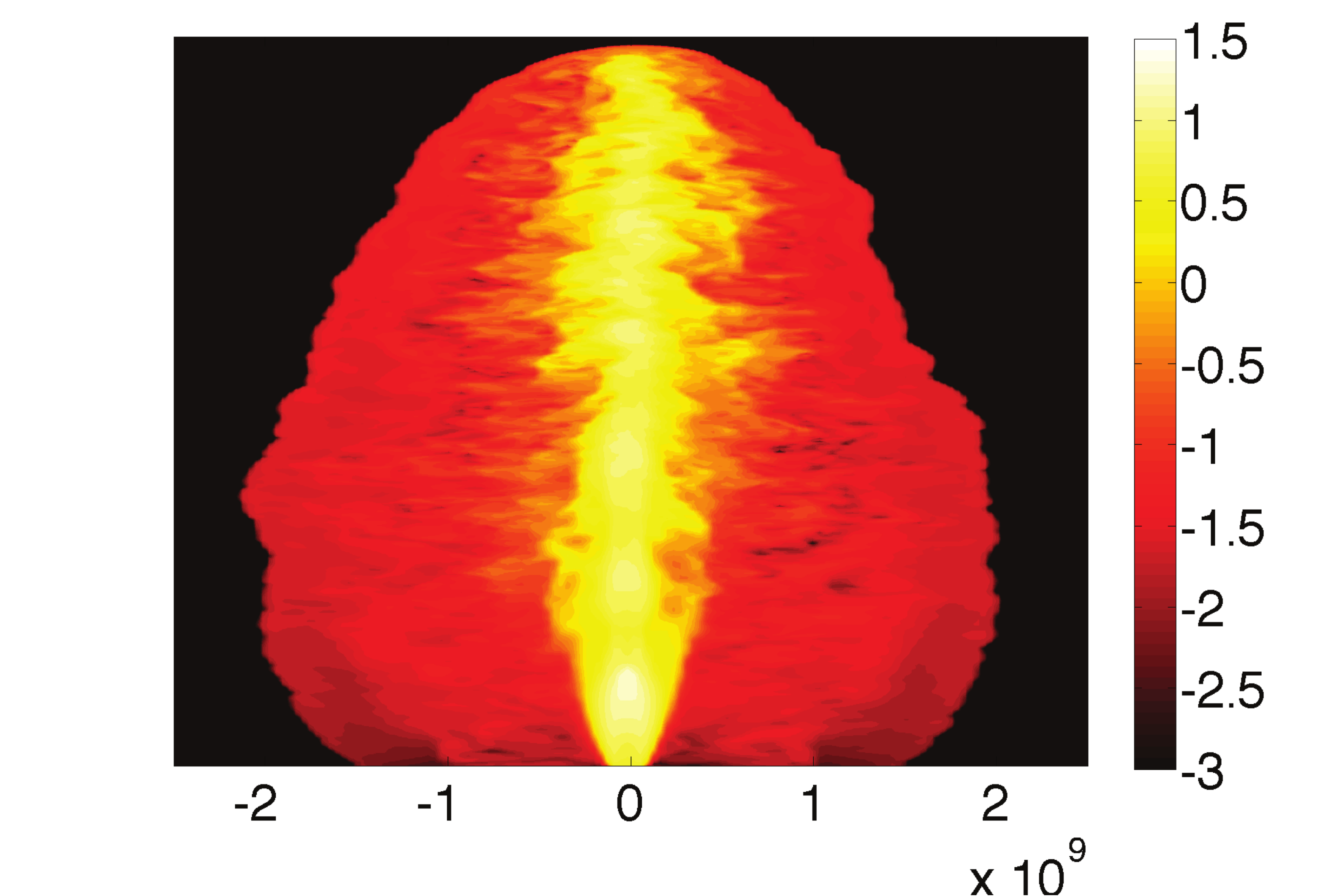}
\end{minipage}

\vspace{5mm}

\begin{minipage}[b]{\columnwidth}
\centering
\includegraphics[width=.45\textwidth]{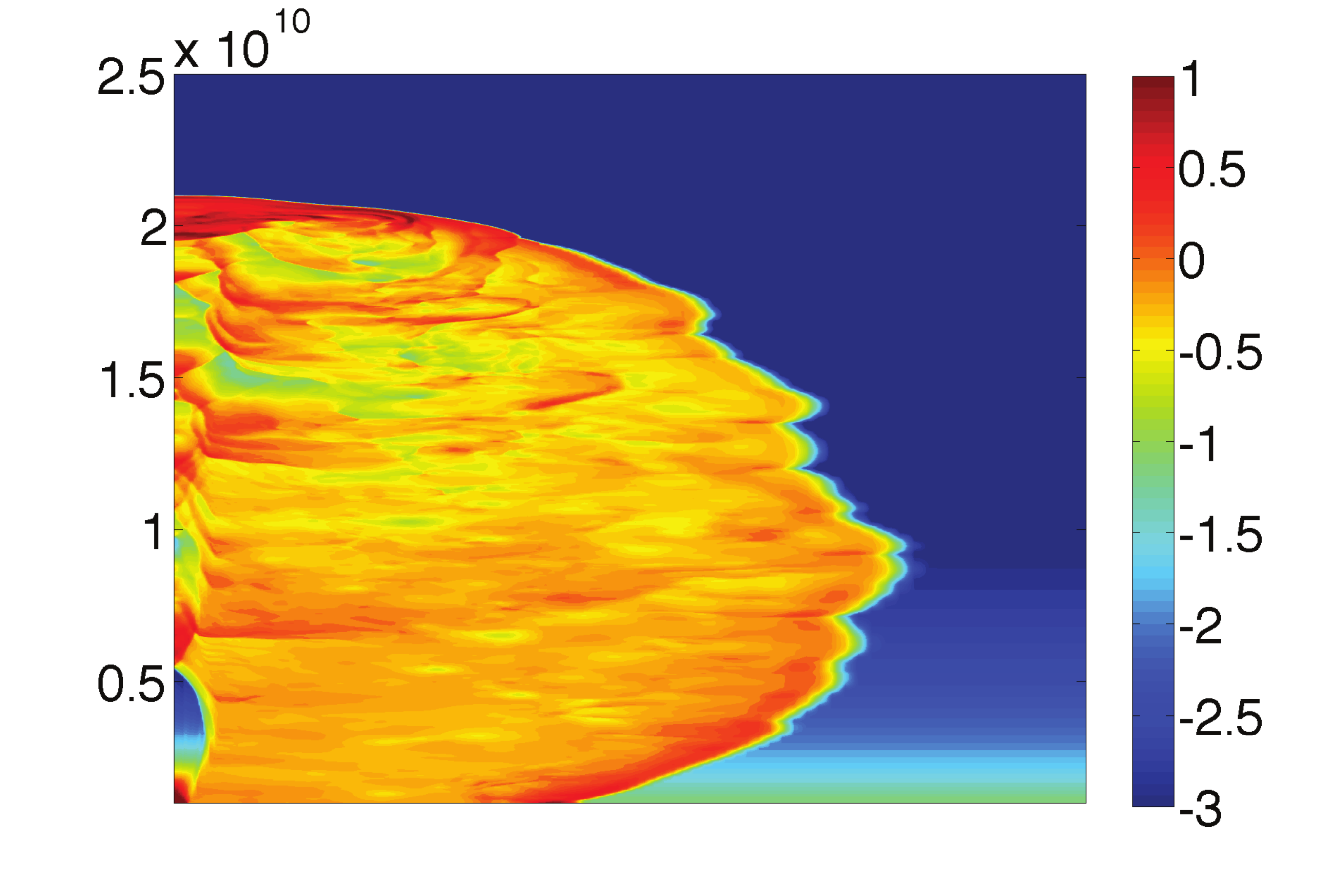}\quad
\includegraphics[width=.45\textwidth]{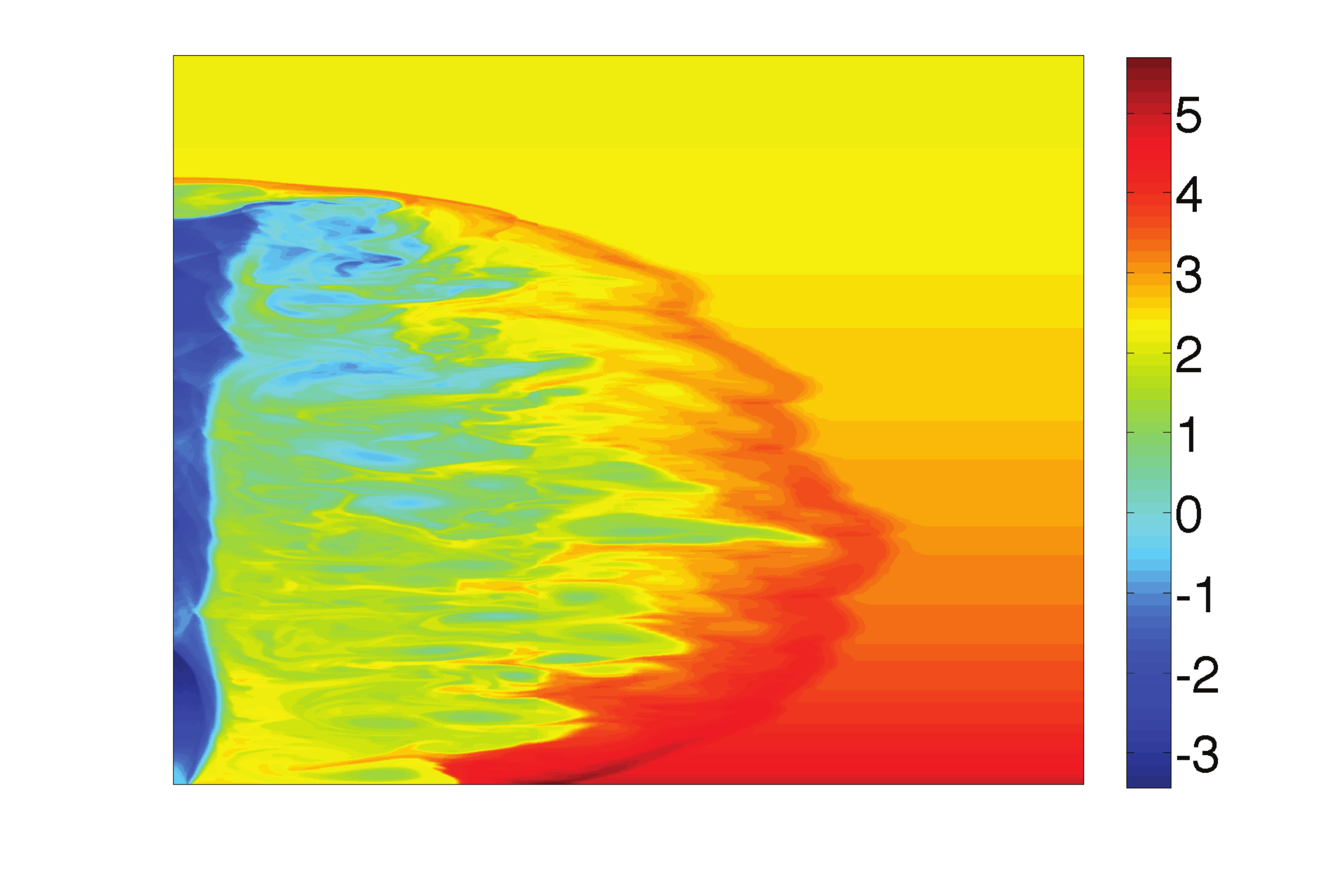}
\end{minipage} \hfill

\begin{minipage}[b]{\columnwidth}
\centering
\includegraphics[width=.45\textwidth]{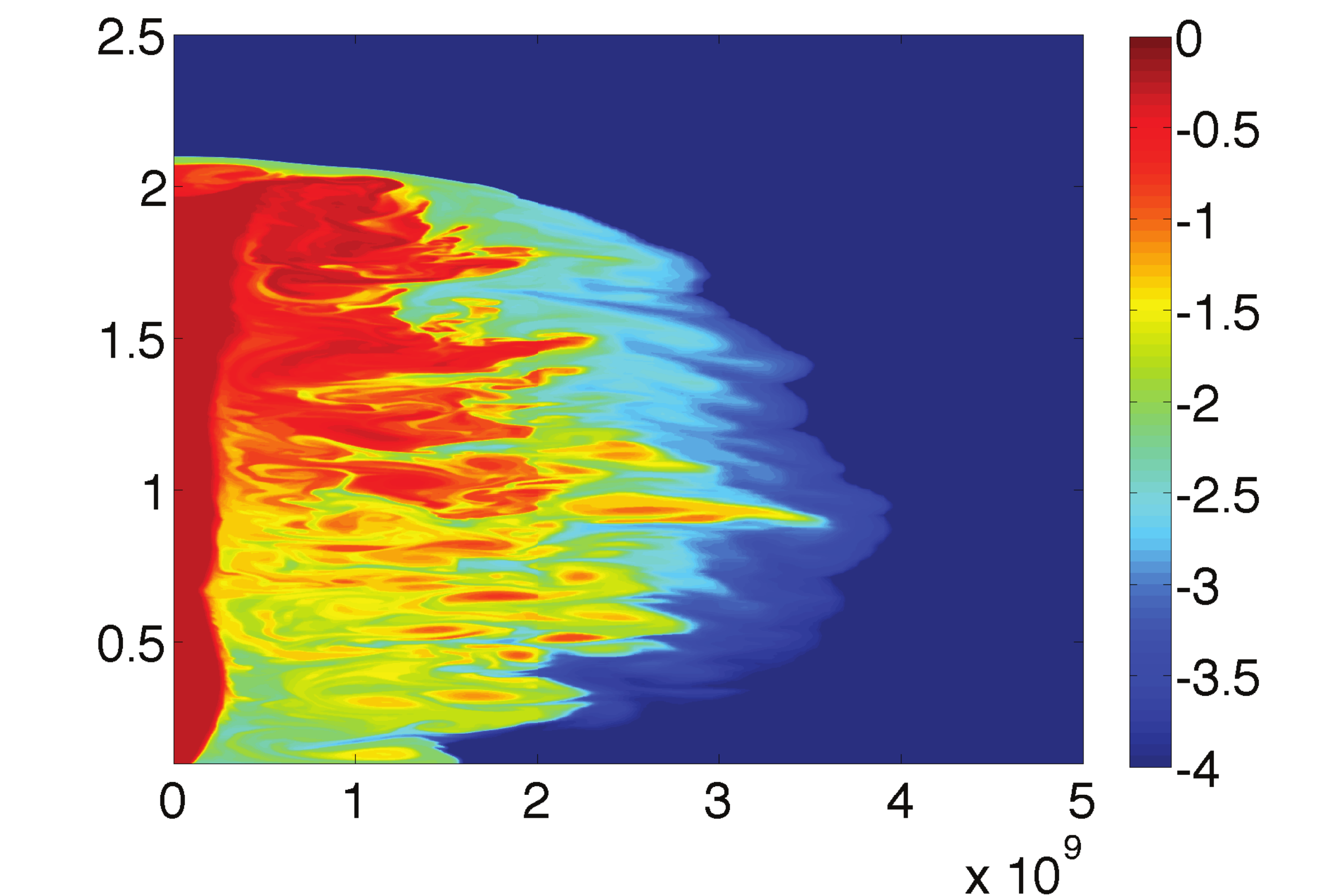}\quad
\includegraphics[width=.45\textwidth]{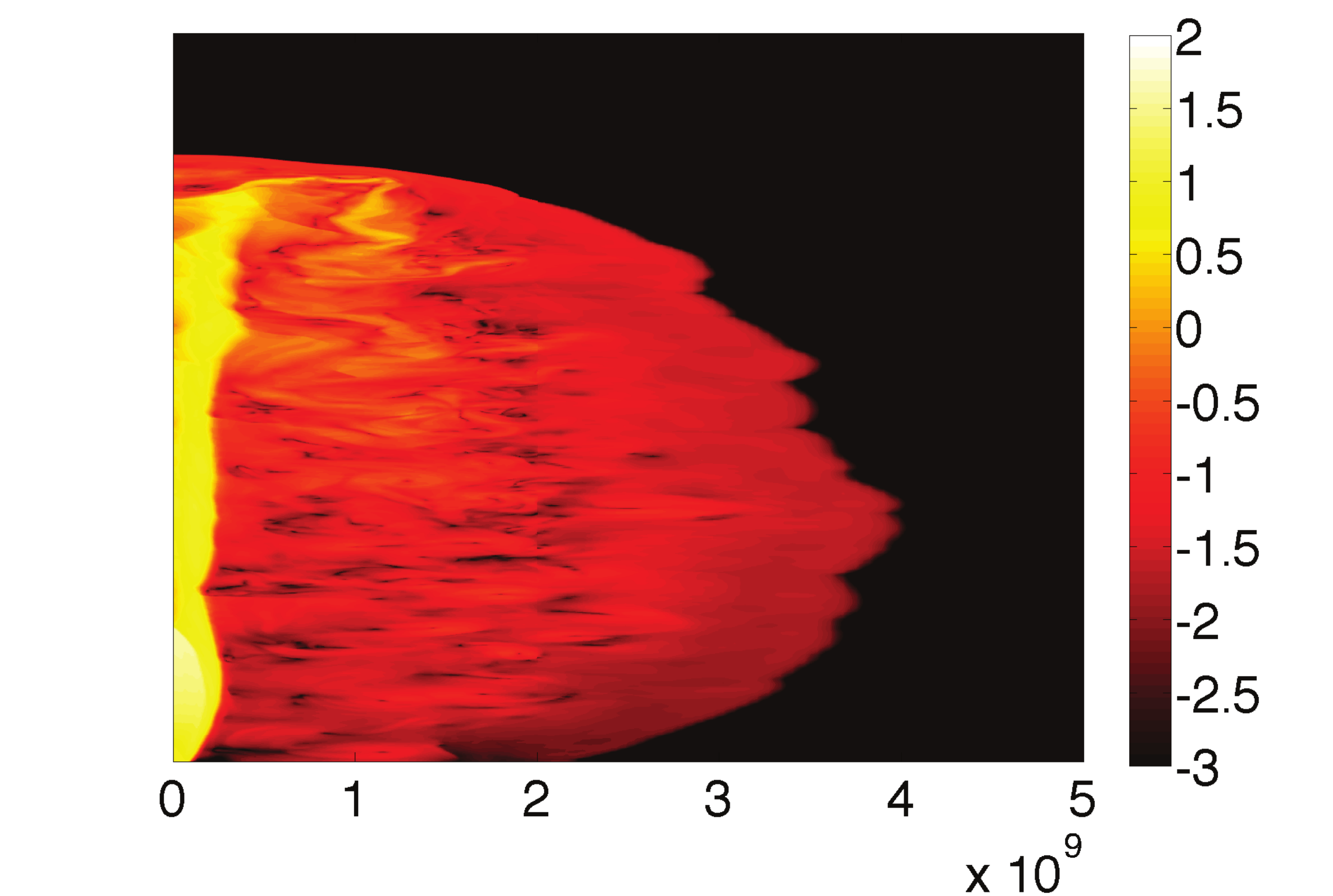}
\end{minipage}
\leavevmode\smash{\makebox[0pt]{\hspace{1em}
\rotatebox[origin=l]{90}{\hspace{8.5em}
{\fontfamily{<familyname>}\selectfont z (cm)}}%
}}
\leavevmode\smash{\makebox[0pt]{\hspace{1em}
\rotatebox[origin=l]{90}{\hspace{27.5em}
{\fontfamily{<familyname>}\selectfont z (cm)}}%
}}

\hspace{12.5em} {\fontfamily{<familyname>}\selectfont r (cm)}

\medskip

\caption{Representative examples of jet-cocoon structure in 2D (simulation 12; lower four panels) and 3D (simulation 3Dc; upper four panels). The figures of the 3D simulations show the x-z plane. The four panels show: top left - pressure ($p/c^2$), top right - density ($\rho$), bottom left - sound velocity ($c_s/(c/\sqrt{3})$), bottom right - four velocity ($\Gamma\beta$). The colormaps are all in log scale cgs. {Note that the scaling of the $r$ and $z$ axes are not 1 to 1, so the shape of the jet and the cocoon is more elongated than appear in the figures (i.e., the jet is narrower). We use an uneven scaling in all the jet figures in the paper.}
\label{fig:4panel}}

\end{figure}
 \begin{figure}
\begin{center}
\includegraphics[width=1\columnwidth]{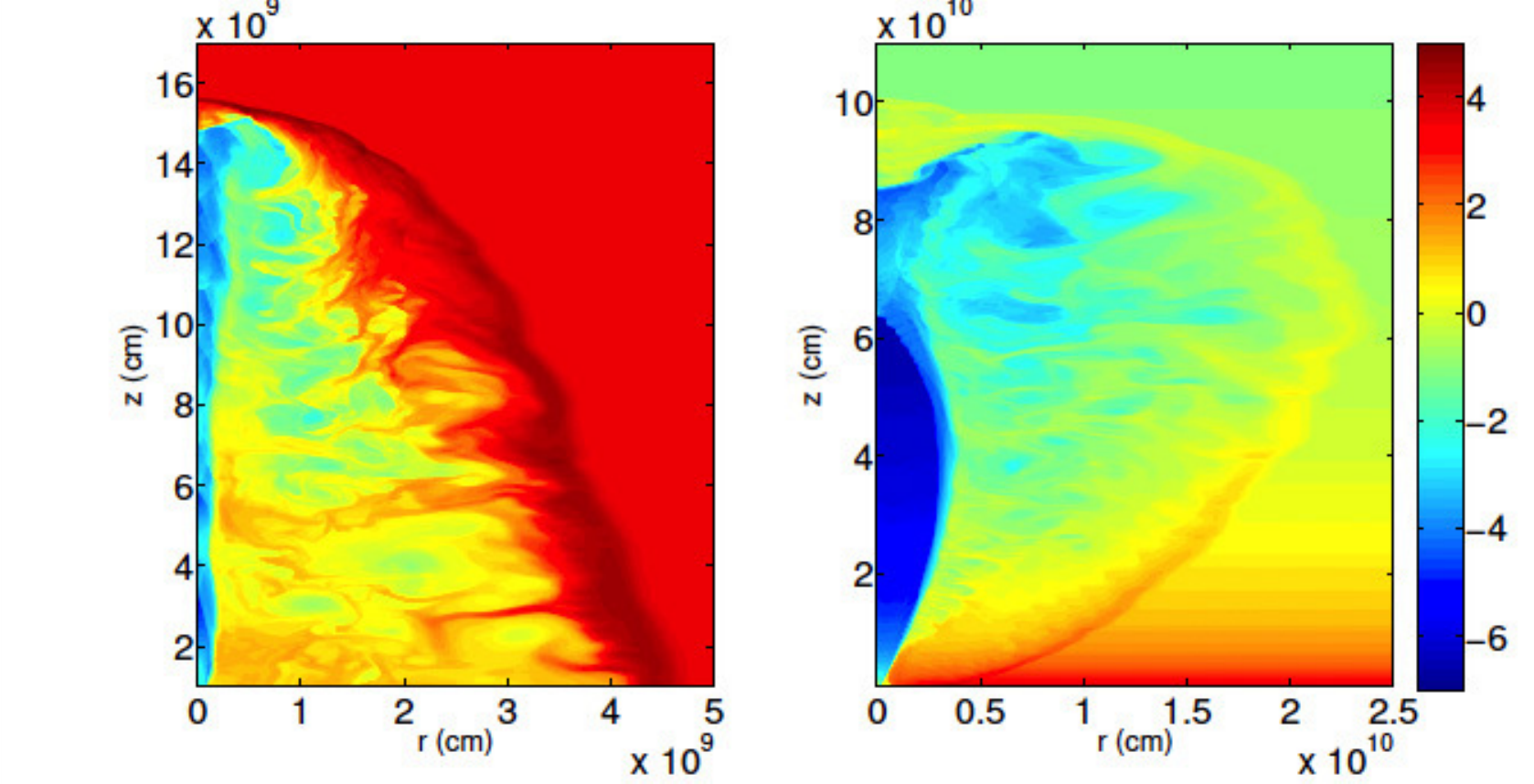}
\caption{Colormap of $\mathrm{log}(\rho)$ for a 2D simulation with external density power-law $\alpha=1$ (left), where the jet decelerate, and  a simulation with $\alpha=2.5$ (right), where the jet accelerate. It demonstrates the different cocoon shapes observed for varying values of $\alpha$.
\label{fig:cocoon}}
\end{center}
\end{figure}


\begin{figure*}

\minipage{0.32\textwidth}
  \includegraphics[width=\linewidth]{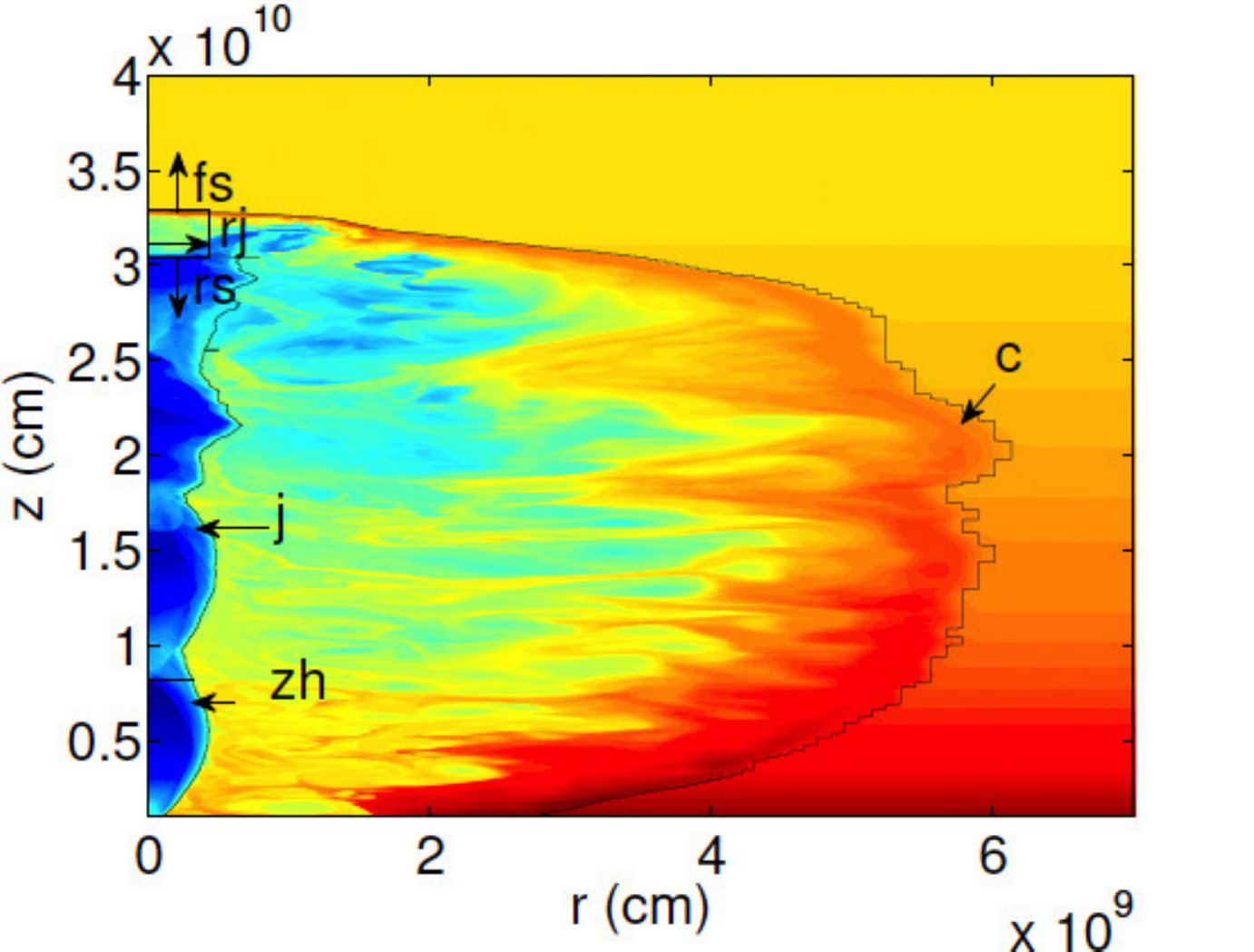}
\endminipage\hfill
\minipage{0.32\textwidth}
  \includegraphics[width=\linewidth]{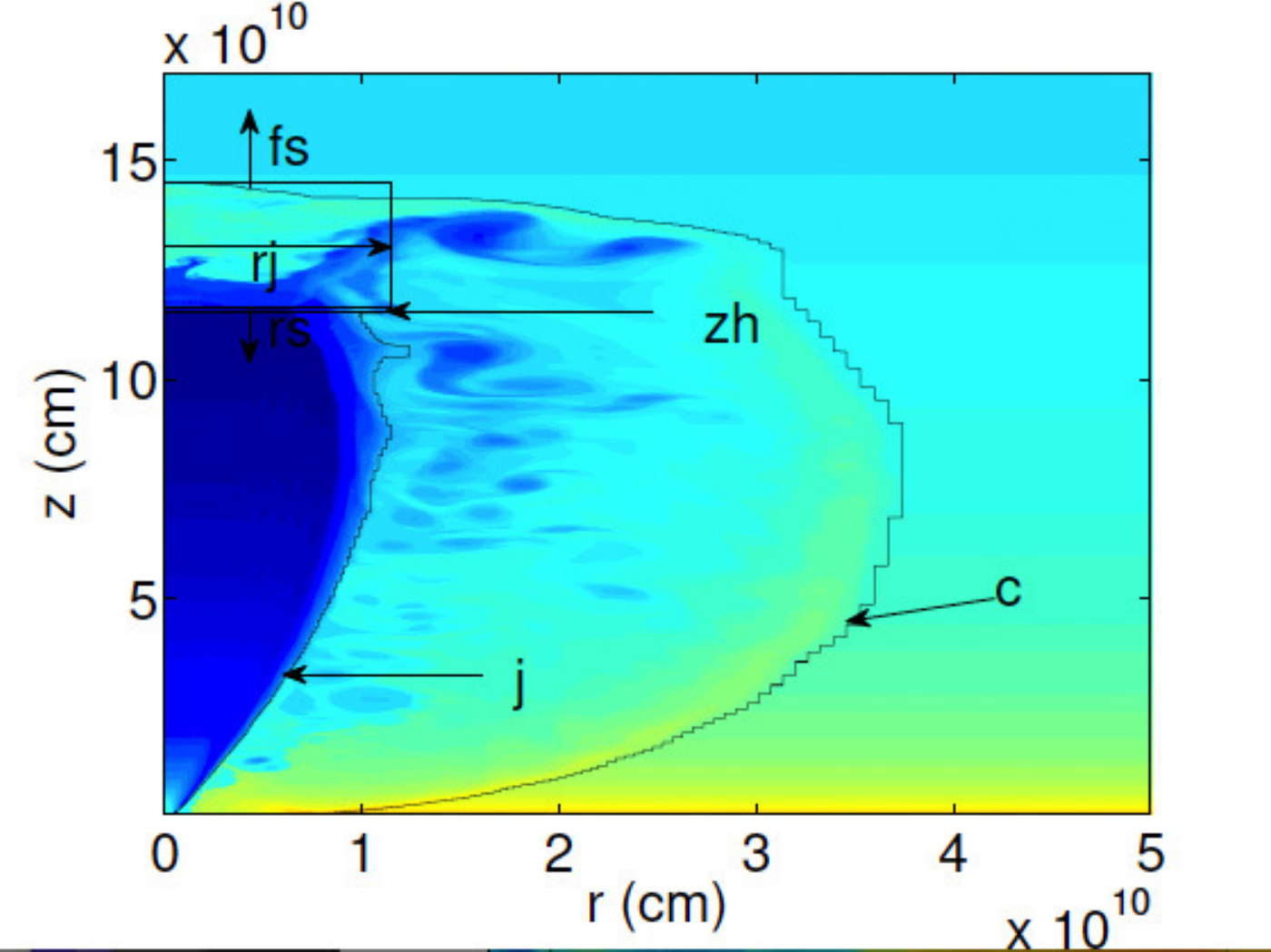}
\endminipage\hfill
\minipage{0.32\textwidth}%
  \includegraphics[width=\linewidth]{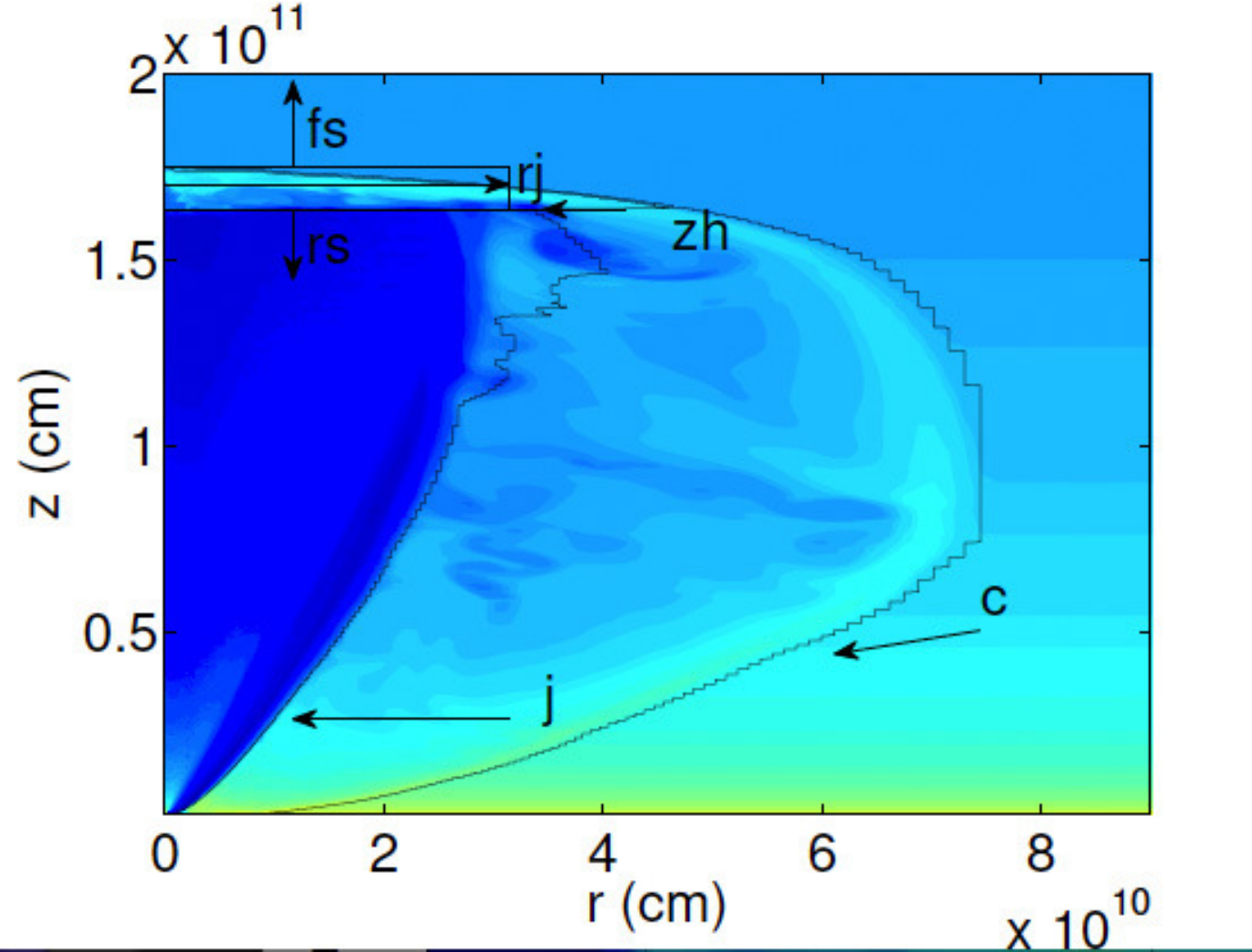}
\endminipage
  \caption{Colormap of $\mathrm{log}(\rho)$ for 3 jet configurations: fully collimated (left), marginally collimated (middle) and uncollimated (right). On each figure we mark the results of our tracing algorithm, which identifies the various regions, and label the forward shock (fs), reverse shock (rs), jet radius (rj), collimation shock height (zh), cocoon boundary (c) and jet boundary (j).}\label{fig:collimation}

\end{figure*}
\subsection{General structure in the collimated regime}
Figure \ref{fig:4panel} shows  a typical structure  of a collimated  jet and  its surrounding cocoon  in 2D and 3D, here with $\alpha=2$ (simulations No. 12 and 3Dc).  The quantities  shown  in the  four panels  are $p$,  $\rho$,
$c_s/(c/\sqrt{3})$ and $\Gamma \beta$, each highlighting  different aspects of the  jet-coocon structure. The
observed structure agrees with the approximated analytic structure in many aspects, but not in
all. In  all our simulations we  see all the main  features of the predicted  structure: unshocked jet,
collimation shock, shocked jet, head with forward/reverse shock and a contact discontinuity, and a cocoon
that   is    separated   to   heavy    shocked   ambient medium    material   and   light    shocked   jet.

Yet, there are some features that are not included in the analytic model. (i) The observed head structure is much more complex than the simplified analytic model. Moreover, the structure of the head in 2D and 3D simulations is very different (see section \ref{sec:3Dvs2D}). Nevertheless, the jet evolution fits the simplified analytic model, to within an order of magnitude, including its estimates of the head cross-section and the resulting head propagation velocity.  (ii) The cocoon is obviously not a perfect cylinder, but it is not too far from that. It looks like an elongated flat headed cone when $\alpha<2$ and an inverted one when $\alpha>2$ (see Figure \ref{fig:cocoon}). For $\alpha=2$  it looks like a roughly symmetric barrel (see Figure \ref{fig:4panel}). These structures have an effect of order unity on the estimated cocoon volume (and thus pressure).  (iii) In all the simulations of collimated jets the shocked jet goes through repeated collimation shocks, as expected \citep[e.g.,][]{sanders1983}. The first shock, which is accounted for  by the analytic model, is stronger than the following, unaccounted, shocks. We do not find that these repeated shocks have a strong effect on the general evolution.  (iv) While the cocoon is separated to low-density (high sound speed)  and high density (low sound speed) regions, a significant mixing between jet material and ambient medium material takes place in the cocoon (see section \ref{sec:mixing}).

\subsection{The transition from the collimated to the uncollimated regime} 
The analytic model provides a criterion for the transition from a collimated to an uncollimated jet (Equation \ref{eq:col}). This criterion is based on the requirement that in the collimated regime the jet head is higher (at larger $z$) than the predicted location of the collimation shock. Figure \ref{fig:collimation} shows jets from three simulations where the transition between the collimated, marginally collimated and uncollimated regimes are seen. The left panel shows simulation 12 where $\L \theta_0^{4/3}(16\Omega/3)^{-2/3}=0.0005$ and the jet is fully collimated. The middle panel shows simulation 16 where $\L \theta_0^{4/3}(16\Omega/3)^{-2/3}=0.3$ and the jet is marginally collimated (i.e., the collimation shock almost converges to the axis). The left panel shows simulation 18 where 
$\L \theta_0^{4/3}(16\Omega/3)^{-2/3}=5$ and the jet is uncollimated.

\subsection{Self similarity in constant $\L$ jets}\label{sec:selfsimilar}
\begin{figure}
\begin{center}
\includegraphics[width=1\columnwidth]{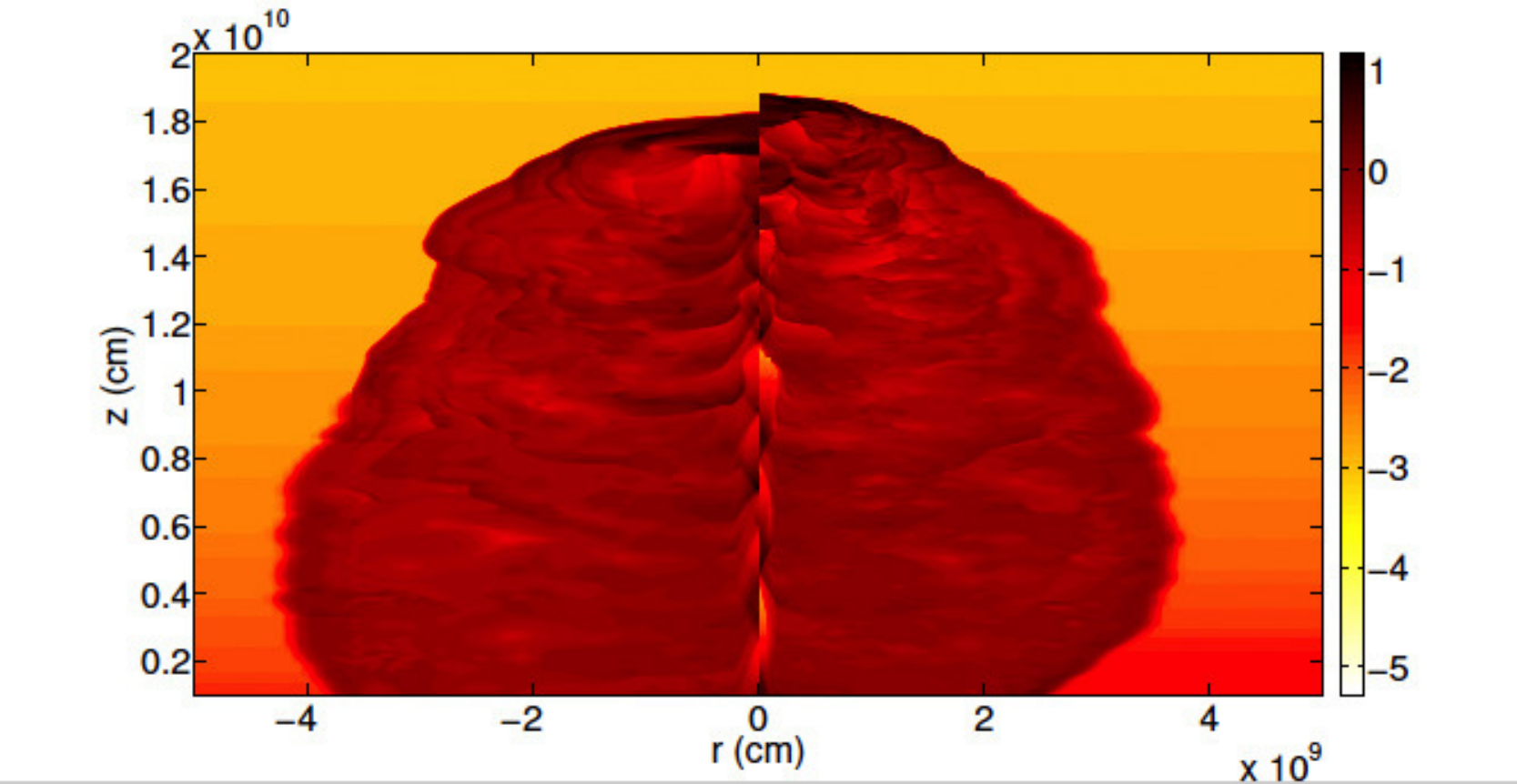}
\caption{Two snapshots of $log(p)$ from simulation 12 at time $t=10\mathrm{sec}$ (left) and $t=3\mathrm{sec}$ (right), with the $10\mathrm{sec}$ snapshot scaled according to the self similarity relations described in the main text. 
\label{fig:scaling}}
\end{center}
\end{figure}

Jets with a constant $\L$ that propagate in a power-law density profile are expected to be self similar where all length and time scales in the shocked region grow as $t$ while mass scales grow as $t^{3-\alpha}$. Thus, the pressure for example scales as $t^{-\alpha}$. Figure \ref{fig:scaling} presents two pressure maps from simulation 12 taken at different times but scaled according to the scaling laws  given above. It is evident that the main properties of the system, such as head and cocoon location, collimation shock height and average pressure all follow the self similar scaling. We have seen this self similarity in all the simulations with constant $\L$. This includes all the simulation with $\alpha=2$ and simulation 29 where $\alpha =1.5$ and $L_j \propto t^{0.5}$ so $\L$ remains constant.

\subsection{Pressure gradients in the cocoon}
\begin{figure}
\begin{center}
\includegraphics[width=1\columnwidth]{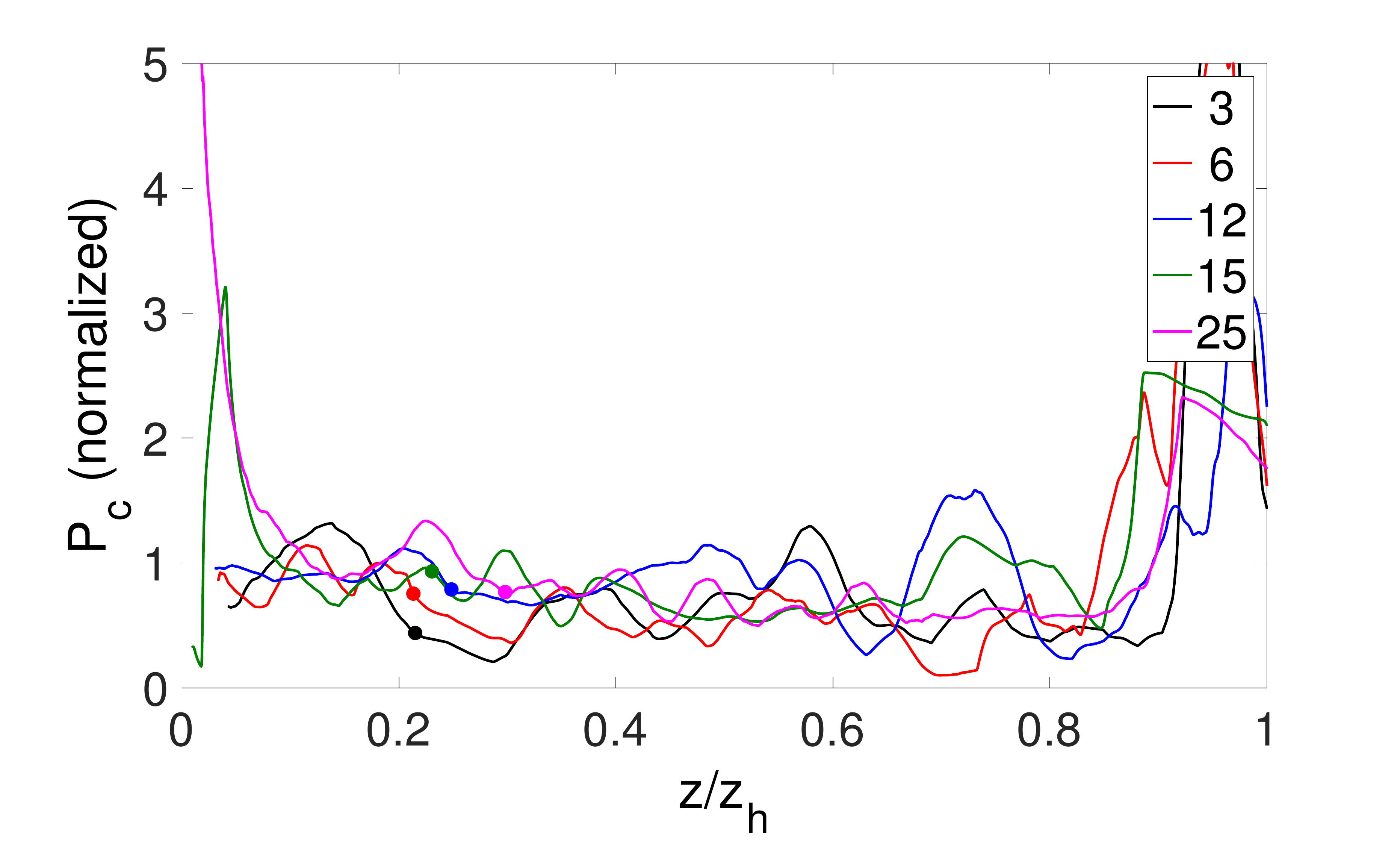}
\caption{The pressure that the cocoon applies on the jet in various simulations (simulation number is given in the legend). $P_c$ at a fixed distance from the jet axis, which is just outside of the jet, is depicted as a function of $z/z_h$. $P_c$ is normalized to the average value between the collimation shock and the head and is smoothed over a scale of $z_h/20$ to avoid local, small scale, fluctuations. The filled circles mark the location where the collimation shock first intersects with the jet axis. The sharp rise in the pressure seen in all simulations around $z/z_h=0.85$ is at the transition to the high pressurized jet head.
\label{fig:prs_grad}}
\end{center}
\end{figure}

The analytic solution approximates the pressure in the cocoon as being uniform. The cocoon pressure on the jet determines the structure of the collimation shock and the jet cross-section between the collimation shock and the head. The roughly cylindrical structure of the jet between the first collimation shock and the head, which is seen in all the simulations where the jet is collimated, implies that the pressure on the jet at different heights is roughly uniform. Figure \ref{fig:prs_grad} depicts the pressure profiles in the cocoon along the $z$ direction near the jet. It presents the profile from several representative simulations of collimated jets, with a range of $\L$ and $\alpha$ values. We also mark the location where the collimation shock converges to the $z$-axis ($\hat{z}$) in each simulation. All curves are normalized to the average pressure between the collimation shock and the jet head. The sharp rise in the pressure around $z/z_h=0.85$ marks the transition to the head. The figure shows that in all simulations the pressure between the collimation shock and the reverse shock is largely uniform with typical fluctuations by a factor of about $2$. These results justify the approximation of a constant pressure in the analytic solution.   

\subsection{Mixing of jet material and energy per particle  in the cocoon}\label{sec:mixing}

The cocoon contains both shocked jet and shocked ambient material that were spilled from the head. It also contains ambient medium material that was shocked as the high pressurized cocoon propagates sideways.  All numerical simulations done so far, find that the shocked jet and shocked ambient material remains largely separated in the cocoon, leading to a light and hot shocked jet material close to the jet axis and heavy colder shocked ambient material far from the axis; yet, some mixing is observed.  The amount of mixing that the shocked jet material goes through is of interest for two reasons. First, it determines the  sound speed in the light hot part of the cocoon, which is responsible for causal connection and pressure equalization throughout the cocoon. Second, it determines to large extent the properties of the cocoon material once the jet is breaking out of its surrounding material and the hot cocoon is free to expand (e.g., as in the case of a GRB jet that breaks out of its stellar progenitor; \citealt{nakar2017}). 

\begin{figure}
\begin{center}
\includegraphics[width=1\columnwidth]{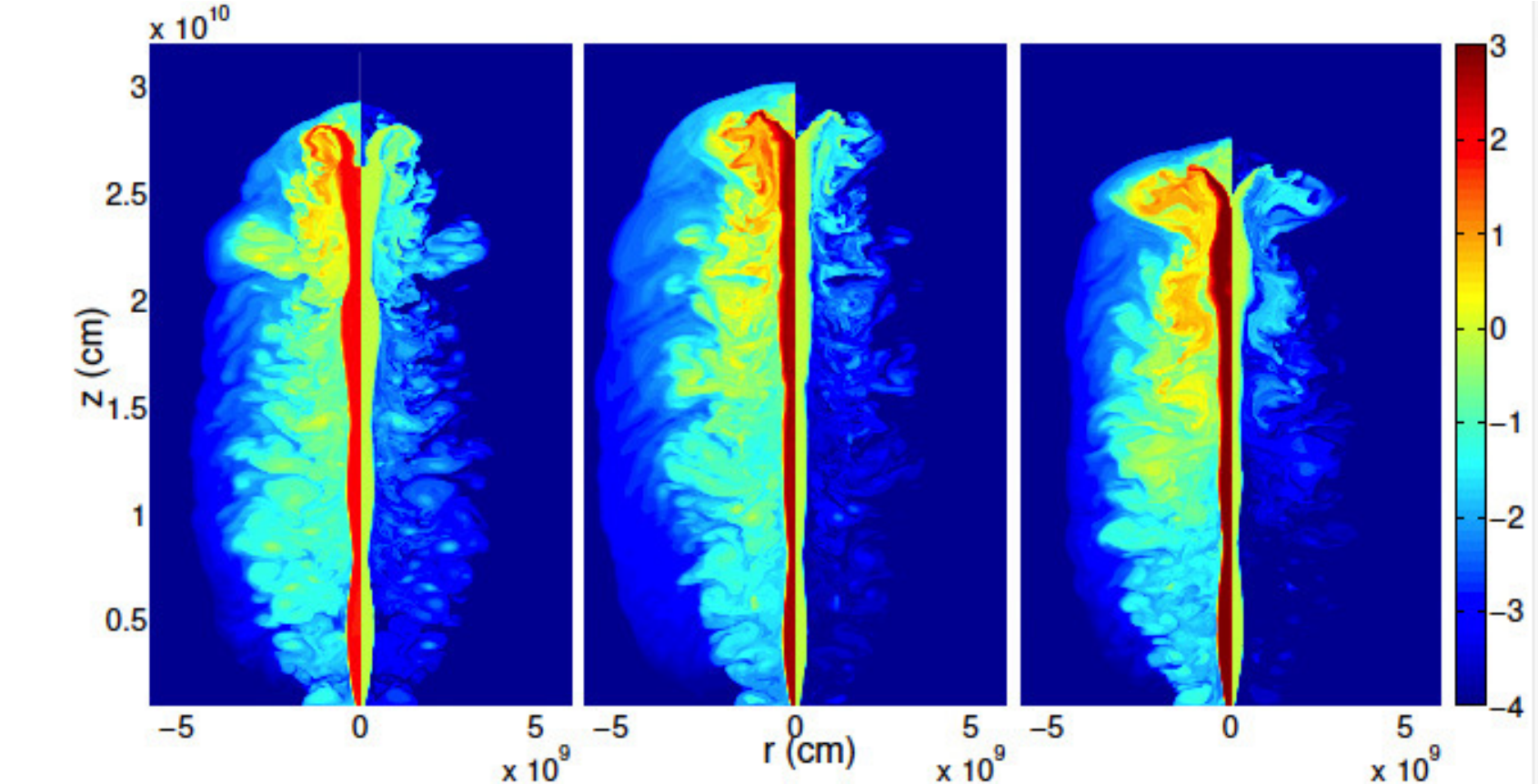}
\caption{To demonstrate the degree of mixing we show three simulations with differing initial specific enthalpy (from left to right: simulations 11, 12 \& 13 ). On the left-side of each figure we show the energy per baryon (logarithmic scale, measured in the lab frame  in units of proton rest mass energy) and on the right-side a passive scalar that is injected with the jet at the nozzle (value of 1 implies a pure jet material). The color scheme is logarithmic and similar in both sides. These figures demonstrate that in 2D mixing does not take place in the jet and that a strong mixing takes place at the head. The shocked-jet material energy per baryon in the cocoon is $\sim 3$ near the head, dropping to $\sim 0.03$ near the base, independent of the initial specific enthalpy of the jet.
\label{fig:mix2d}}
\end{center}
\end{figure}

A full characterization of the mixing, its origin and its dependance on the physical parameters, is beyond the scope of this paper. Here we give only a brief description of the mixing observed in our various simulations. In all simulations, 2D and 3D, we observe a partial, yet significant, mixing. We repeated several simulations (both in 2D and 3D) with {resolutions that are lower by 1.5 (i.e. grid cells size bigger by a factor of 1.5) and using a tvdlf solver,} and obtained similar mixing patterns and levels, suggesting that the origin of the mixing in the simulations is dominated by physical and not numerical processes.
In all our 2D simulations we see a similar mixing which is demonstrated in Figure \ref{fig:mix2d}. This figure shows the density of passive trace particles of jet material and the energy per baryon in three simulations which are similar in all parameters except for the initial energy per baryon in the jet (simulations 11-13).  In all simulations the jet itself remains unmixed up to the head where a strong mixing takes place mostly via oblique shocks. The mixing at the head reduces the energy per baryon to be of order unity (in units of the particles rest-mass energy), regardless of the initial energy per baryon in the jet. Namely, the mixing always brings the temperature of the shocked jet gas at the upper part of the cocoon to be mildly relativistic (i.e., the mixing is inversely proportional to the initial specific enthalpy). Later, the mixing continues within the cocoon, reducing the energy per baryon at the the lower parts of the cocoon to $\sim 3 \times 10^{-2} {\rm ~m_p c^2}$ (i.e., sound speed of $\sim 0.1$c).

\begin{figure}
\begin{center}
\includegraphics[width=0.8\columnwidth]{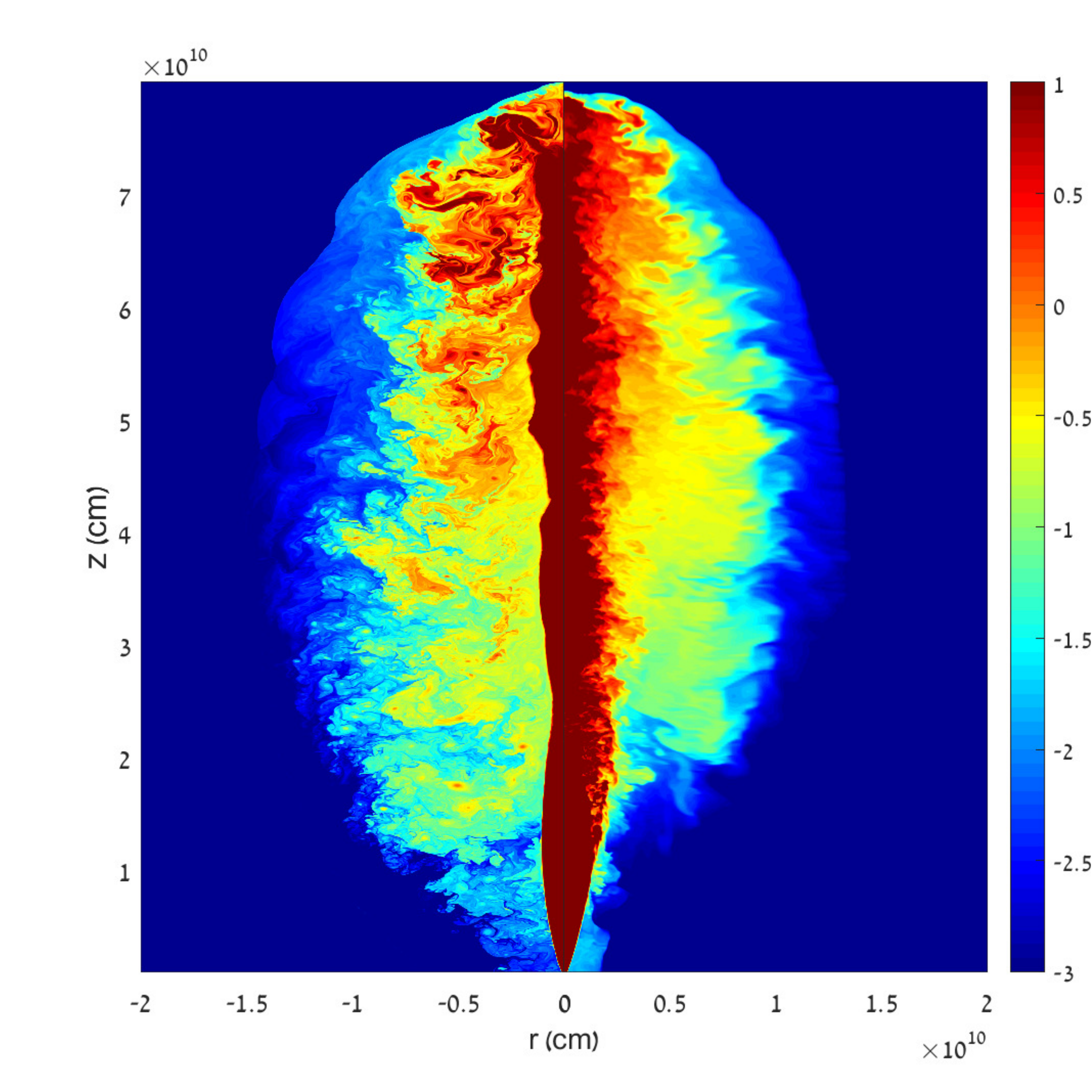}
\caption{A comparison of the energy per baryon in 2D (left side) and 3D (right side) simulations. It shows that the mixing levels in both simulation types are similar, although at least some of the mixing takes place at different locations.
\label{fig:mix3d2d}
}
\end{center}
\end{figure}

3D simulations also exhibit a significant mixing at levels that are roughly similar to 2D  (see Figure \ref{fig:mix3d2d}). Similarly to 2D simulations the gas in the shocked jet cocoon is mildly relativistic  near the head and mixing is progressively higher at lower parts of the cocoon. However, when looking at details there are significant differences between the 2D and 3D mixing sites, suggesting a different origin for the mixing. Thus, in order to understand the mixing properly it should be investigated in details in 3D simulations. We defer such study to a future work.

 \begin{figure}
\begin{center}
\includegraphics[width=0.8\columnwidth]{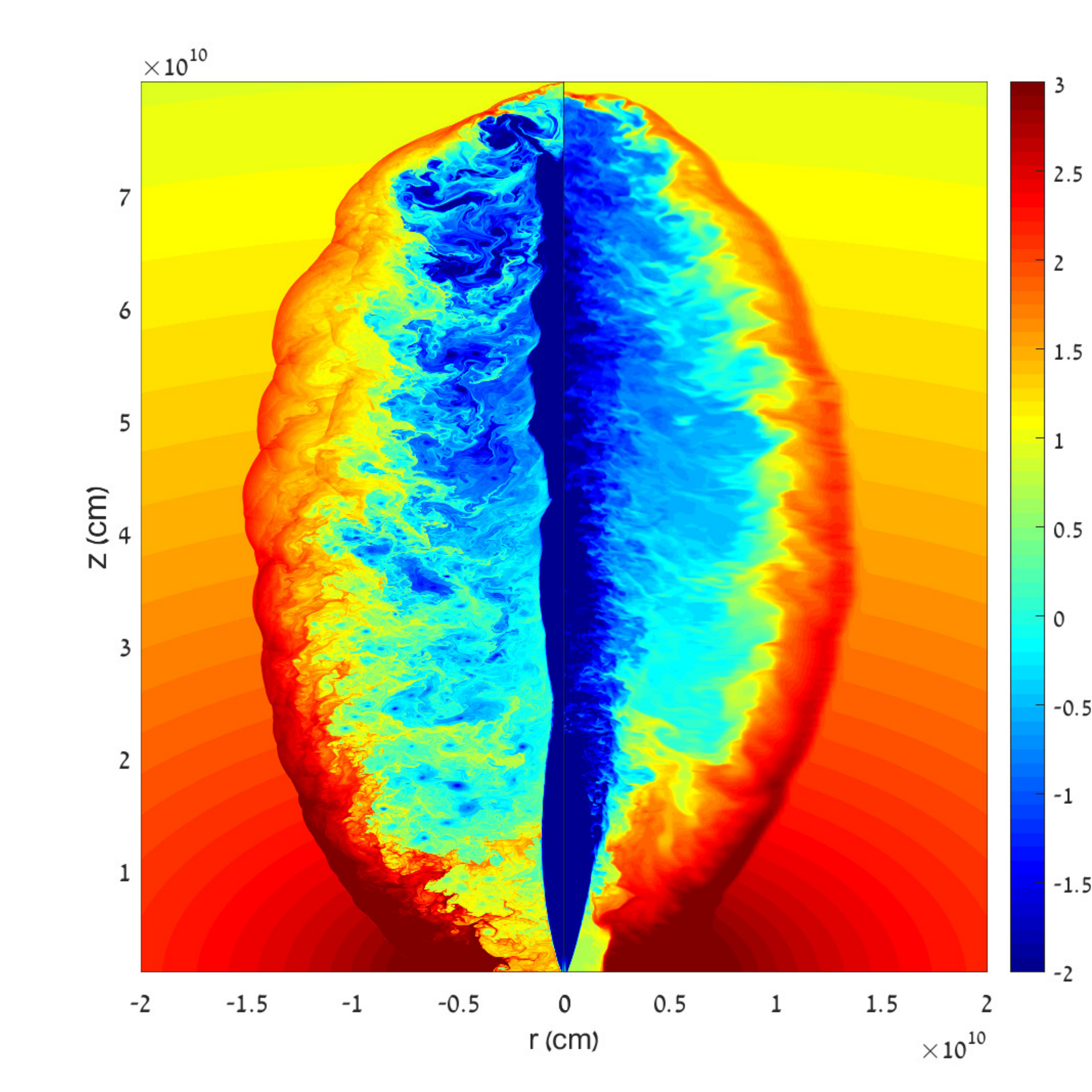}
\caption{A density colormap of $log_{10}(\rho)$ of 2D simulation (left side) and the corresponding 3D simulation (right side). This comparison shows the similar general structure as well as the differences in details such as the head structure.
\label{fig:3drho}
}
\end{center}
\end{figure}



\subsection{Comparison between 2D and 3D simulations}\label{sec:3Dvs2D}
We carried out half a dozen 3D simulations (table \ref{tab:sim3d}) that correspond to 2D simulations with the same parameters. We found many similarities, but also important differences. The general morphology of a jet, collimation shocks and a cocoon is similar (Figure \ref{fig:3drho}). The jet is collimated roughly at the same height in both simulations and the jet remains rather cylindrical up to the head. The cocoon has a similar shape and is separated to light shocked jet and heavy shocked ambient material. However, there are also morphological differences. Most noticeable is the structure of the head. In 2D the head structure is strongly affected by the imposed axial symmetry, developing a plug that `seats' on the head of the jet (see also \citealt{macfadyen2001,zhang2004,lopez-camara2013}). Some of the jet material is deflected sideways by the plug dissipating its energy via oblique shocks that looks in 2D like a pair of antennae. This results in a head that is wider than the jet and thus at a slower propagation (see equation \ref{eq:Ltild}). In 3D the head looks very different. {Its structure is much less regular and the plug does not play a decisive role in the jet's head evolution (see also \citealt{lopez-camara2013,ito2015}).} Instead the head losses its cylindrical shape, get deformed and its cross section increases somewhat. The resulting head velocities in the corresponding 2D and 3D simulations we carried out are comparable to within a factor of order unity. Similarities and differences between 2D and 3D simulations are also seen in the mixing between jet and ambient material in the cocoon (see section \ref{sec:mixing}).

\subsection{The  power-law indices of the temporal evolution}
\begin{figure}
\begin{center}
\includegraphics[width=1\columnwidth]{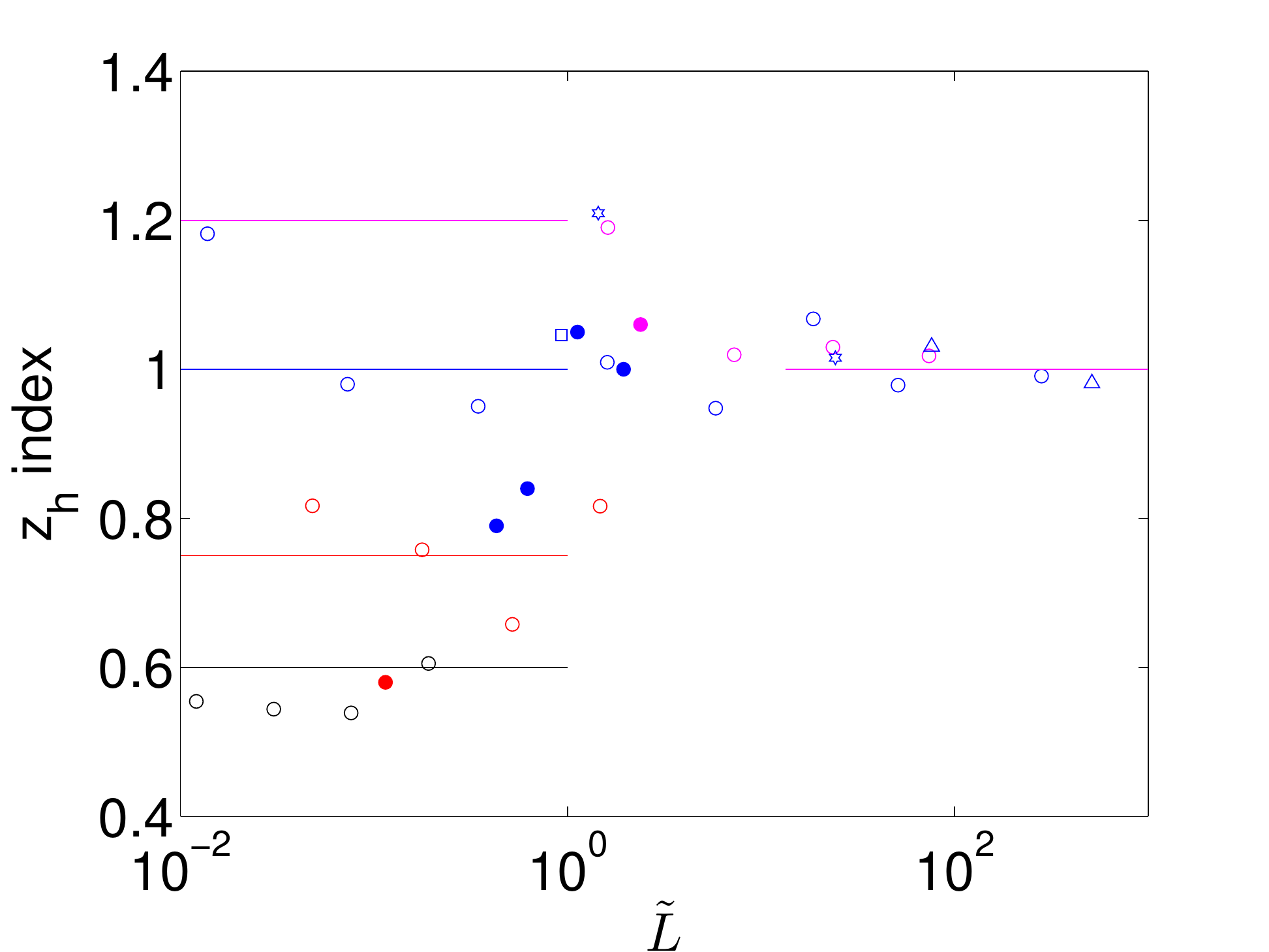}
\caption{The temporal power-law index of the head location as function of $\L$. 2D simulations are marked by {\it open symbols}, 3D simulations by {\it filled symbols} and the analytic model by {\it solid lines}. The symbol represent $\theta_0=0.14$ rad ({\it circles}), $\theta_0=0.04$ rad ({\it triangles}), $\theta_0=0.07$ rad ({\it hexagrams}), and $\theta_0=0.18$ rad ({\it square}). The color coding represents $\alpha=0$ ({\it black}), $\alpha=1$ ({\it red}), $\alpha=2$ ({\it blue}) and $\alpha=2.5$ ({\it magenta}). 
\label{fig:zhscale}}
\end{center}
\end{figure}

\begin{figure}
\begin{center}
\includegraphics[width=1\columnwidth]{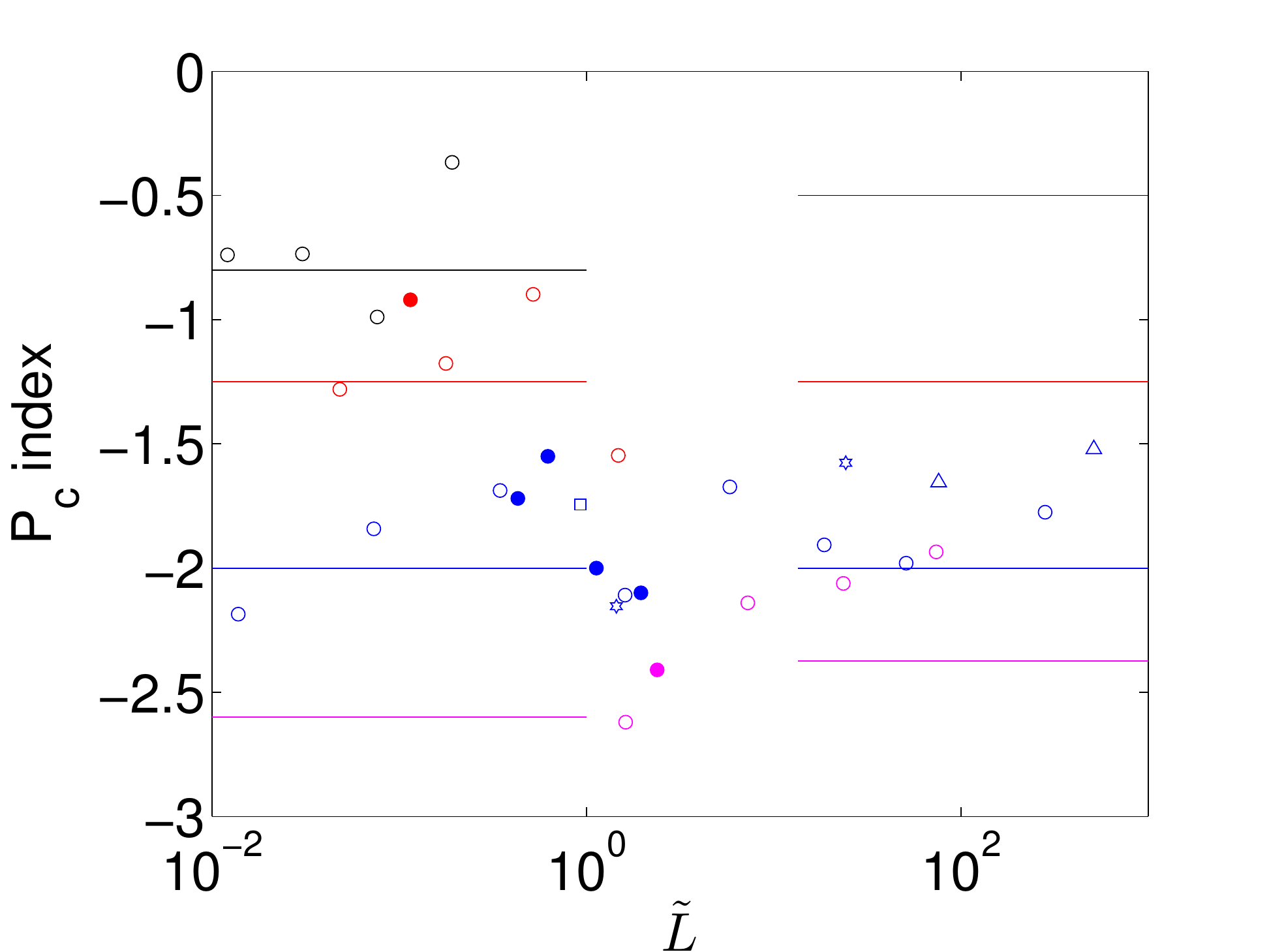}
\caption{The temporal power-law index of the average cocoon pressure as function of $\L$. Markers and color coding are the same as in Figure \ref{fig:zhscale}. 
\label{fig:pcscale}}
\end{center}
\end{figure}

In case of a power-law density profile and constant jet parameters ($L_j$ and $\theta_0$) the analytic solution predicts that all the various quantities of the solution would evolve as power-laws in time in both the Newtonian and relativistic regimes. In appendix \ref{app:analyticModel} we provide the full analytic solutions for the two regimes.
The solutions show that for $\alpha=2$ the power-law indices of the Newtonian and the relativistic regimes are the same (and they all follow the self-similar solution discussed above), while for $\alpha \neq 2$ the indices of the two regimes are different. In figures \ref{fig:zhscale} and \ref{fig:pcscale}  we present the temporal power-law indices of $z_h$ and $P_c$ as a function of $\L$ for all the simulations we carried out. The figures show that the power-law indices measured in the numerical simulations are in good agreement with the analytic predictions, especially those of the head location\footnote{Power-law solutions are predicted also if the jet parameters evolve as power-laws in time. In figures \ref{fig:zhscale}-\ref{fig:pcscale} we present only the cases of constant jet parameters. The single case that we examined of a varying $L_j$ (but constant $\L$; simulation 29)  also followed the power-law evolution predicted by the analytic model, as discussed in the self-similarity section (\ref{sec:selfsimilar}).}.



\section{Calibrated semi-analytic and fully analytic models}\label{sec:calibration}
While the power-law indices of the analytic solution are expected, and found, to be accurate, the numerical coefficients of the analytic model are expected to be accurate only to within an order of magnitude. We use our set of simulations in order to test the accuracy of the numerical coefficients of the B11 model and calibrate it if needed. Since most of the approximations used in the analytic model are on the collimation process, we expect that in the uncollimated regime, where most of the jet flows ballistically from the base till the head,  the model will approach the numerical result. Therefore we expect that calibration may be needed only for the  collimated regime.

\subsection{Semi-analytic model}
\begin{figure}
\begin{center}
\includegraphics[width=1\columnwidth]{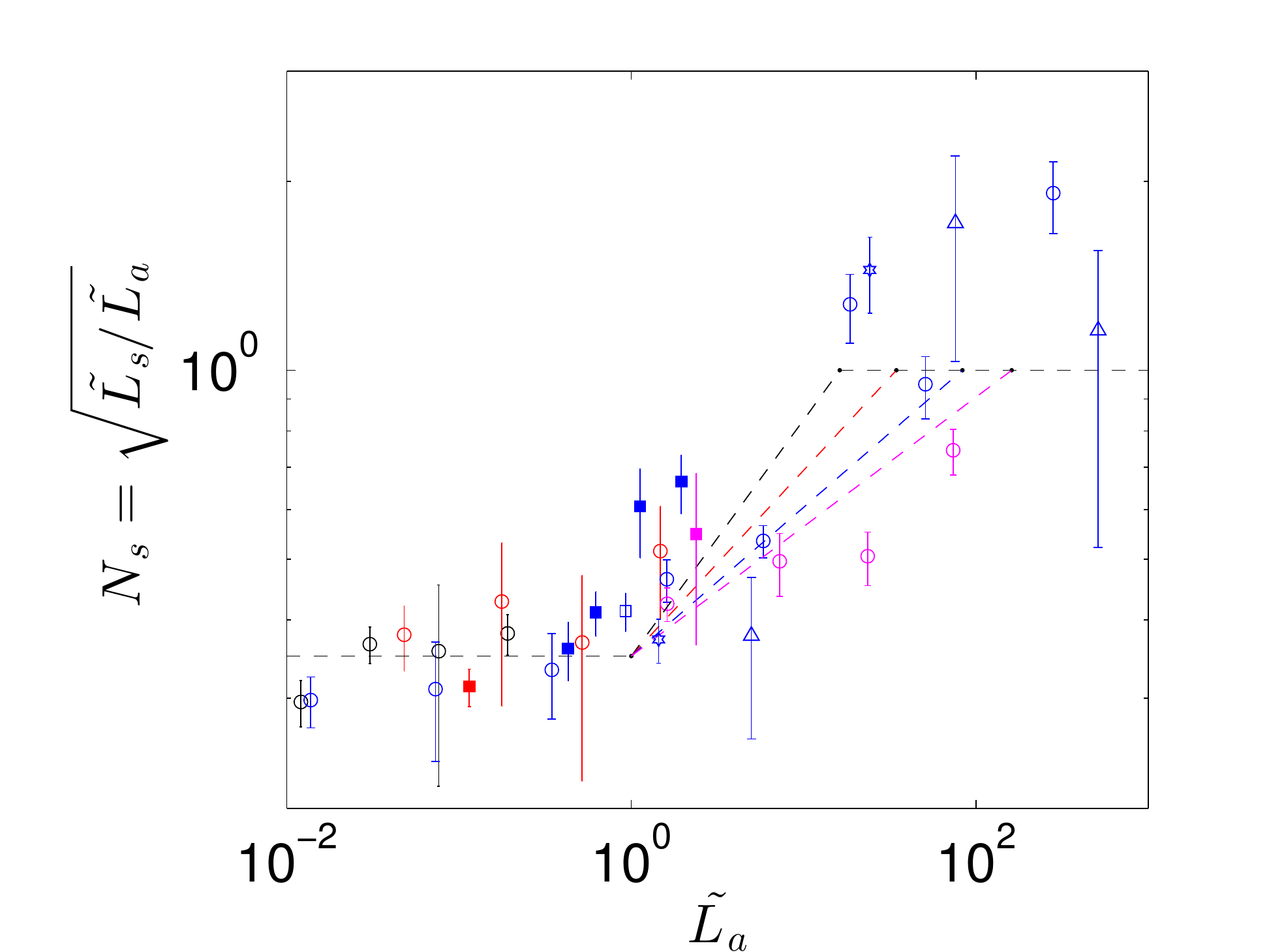}
\caption{The ratio of the numerical value $\sqrt{\L_s}$ and the analytic prediction $\sqrt{\L_a}$ as a function of $\L_a$. In the Newtonian regime this ratio is approximately the head velocity ratio. The markers and color coding are the same as in Figure \ref{fig:zhscale}. The data points represent the average value over the simulation duration after it becomes stable (see appendix \ref{app:convergence}). The error bars represent the standard deviation over the same time, which reflects the fluctuations in the head velocity during the simulation. The lines are our fit to the results which in the transition from the Newtonian collimated regime to the relativistic uncollimated regime depends on $\alpha$.
\label{fig:bh1}}
\end{center}
\end{figure}

In the semi-analytic model the jet properties are integrated over time. Namely  at any time step the head location is given and so is the density at the 
head location.  We can therefore use equations \ref{eq:Ltcol} and \ref{eq:Ltuncol}, where $\rho_a$ is used explicitly, in order to evaluate the analytic 
value of $\L$. Thus, in order to find the difference in the coefficients between the semi-analytic model and the numerical simulations we use equations 
\ref{eq:Ltcol} and \ref{eq:Ltuncol} to calculate the analytic estimate for $\L$, denoted $\L_{a}$, for a given set of jet and medium parameters at a 
given time $t$. The location of the head at time $t$ and thus the density at the head location, $\rho_a(z_h)$ is taken from the simulation and $\Omega$ 
is calculated using equation \ref{eq:Omega}. In order to calculate the numerical value of $\L$, denoted $\L_s$, we measure the velocity of the head and use 
equation \ref{eq:hvel} to find $\L_s$. Figure \ref{fig:bh1} shows the ratio $\sqrt{\L_s/\L_a}$, which in the Newtonian regime is the head velocities ratio. It 
shows that for $\L \lesssim 1$ the head velocity in 2D simulations is slower by about a factor of $2.5-3$ than the analytic prediction of B11. This value 
is similar to the one found by \cite{mizuta2013}. This ratio is largely independent of jet and medium parameters (as long as $\L \lesssim 1$). In the 
uncollimated regime the analytic and numerical values of $\L$ agree, as expected. Linear extrapolation (in $log(\L_a)$) between $\L_a=1$ and $\L_a = 
\theta_0^{-4/3} (16\Omega/3)^{2/3}$ (the transition to uncollimated jet) provides a reasonable approximation to the ratio $\sqrt{\L_s/\L_a}$ for 2D 
simulations. Our 3D simulations cover a smaller range of $\L$ than the 2D simulation set. Yet for $\L<1$, where we expect a significant difference 
between the model and the simulations, we obtain comparable head velocities in 2D and 3D. We therefore obtain that the calibration factor, $N_s \equiv 
\sqrt{\L_s/\L_a}$, can be approximated as $N_s=0.35$ in the Newtonian regime, $N_s =1$ in the uncollimated relativistic regime, and a linear 
interpolation in log-space between the two regimes. The calibrated model is then:
\begin{equation}
	\L_s = N_s^2 \L_a \\
\end{equation}

where $\L_a$ is given in equations \ref{eq:Ltcol} and \ref{eq:Ltuncol}, 
\begin{equation}
	N_s= \left \{ \begin{array}{ll}
         0.35 &  \L_a \leq 1 \\
         0.35 \L_a^{\frac{0.46}{\log_{10}{\L_{col}}}} & 1 \leq \L_a \leq \L_{col} \\
		 1 & \L_{col} \leq \L_a
		 \end{array} 
		 \right.
\end{equation}
and $\L_{col}=\theta_0^{-4/3}(16\Omega/3)^{2/3}$ marks the transition from a collimated to an uncollimated jet.
{In the Newtonian regime ($ \L \ll 1 $) the calibrated head velocity  is therefore $ \beta_h = N_s \L_a^{1/2} $.}
Note that $N_s$ is the calibration coefficient for the semi-analytic model where the head location at a given time is known and $\L_a$ is found using the correct value of $\rho_a$ explicitly. In \url{http://www.astro.tau.ac.il/~ore/propagation.html} we provide an applet that calculates a jet propagation semi-analytically using the calibration factor $N_s$. The algorithm used by the applet is described in appendix \ref{app:SemiAnalyticModel}. 


\subsection{Analytic model in the Newtonian regime}
In case of a Newtonian head velocity ($\L \ll 1$) and a power-law external density we can obtain a full analytic solution. In that case $\rho_a(t)$ can be found analytically and it is more convenient to plug it into equation \ref{eq:Ltcol}, so all quantities depend explicitly only on $t$. As can be seen in Figure \ref{fig:bh1} the head velocities found in the numerical simulations are significantly lower than the analytic predictions in this regime and therefore a calibration is needed. However the calibration coefficient of the head velocity in the fully analytic model is not $N_s$. The reason is that in the uncalibrated semi-analytic solution the density in equation \ref{eq:Ltcol} is the density at the actual head location, while in the uncalibrated analytic solution (where the dependence on $\rho_a$ is implicit) $\rho_a(t)$ is taken at a location where the head would have been at time $t$ if the uncalibrated analytic model would have been correct. In practice, from equations \ref{eq:Ltcol} and \ref{eq:hvel} we find that for $\L \ll 1$ the head velocity in the corrected semi-analytic model is $\beta_h \propto N_s(\rho_a t^2)^{-1/5}$. To eliminate the explicit dependence on $\rho_a$ we plug in $\rho_a \propto (\beta_h t)^{-\alpha}$, obtaining $\beta_h \propto N_s^{\frac{5}{(5-\alpha)}} t^{\frac{(\alpha-2)}{5-\alpha}}$. In appendix \ref{app:analyticModel} we provide the analytic equations for a general value of $\alpha$, here we give only the calibrated head velocity for $\alpha=2$ (i.e., $\rho_a=A_\rho z^{-2}$) and $N_s=0.35$: 
\begin{equation}
	\beta_h = 0.2 \left( \frac{L_{j,50}}{A_{\rho,22}\theta_{0,10^o}^4}\right)^{\frac{1}{3}} 
\end{equation}

\section{A Simulation of jet propagation in a stellar envelope}\label{sec:star}
One application of the theory we consider here is long GRBs where a relativistic jet propagates within a stellar envelope. The density profile of a star is not a power-law and for typical parameters the head velocity is mildly relativistic, so the fully analytic solution is not applicable. We carried out a detailed 3D simulation of a propagation of such  a jet in order to compare the results of our semi-analytic model and to present a more realistic case with a direct implication to an astrophysical phenomenon.

The stellar envelope is assumed to be static, non-rotating and with no magnetic fields. Its radius is $ R{_\star} = 10^{11}\cm $, and its density profile is given by
	\begin{equation}
	\rho(r) = A_\rho r^{-2}\Bigg(\frac{R_\star-r}{R_\star}\Bigg)^3~,
	\end{equation}
where $ A_\rho$ is the normalization, set to be $ 6.4\times 10^{22}~\rm{gr}/\rm{cm} $ such that the total stellar mass is $ M_{\star} = 10\msun $. Note that $r$ is the distance from the center, not to be confused with the cylindrical coordinate we used in all other simulations. The jet is injected at a height $ z_{beg} = 2.5 \times 10^9\cm $. The cylindrical nozzle radius is $2 \times 10^8\cm$ with a homogeneous distribution across its injected planar cross section. The jet opening angle is $0.14$ rad ($ \Gamma_{j,0} = 5 $) and it is injected with a specific enthalpy $ h_{j,0} = 108 $. The total two-sided luminosity of the bi-polar jet is $ L_{j,tot} = 2.1 \times 10^{50}~\rm{erg/s} $ corresponding to an isotropic equivalent luminosity of $ 2.1\times 10^{52}~\rm{erg/s}$.

{The 3D Cartesian grid has five patches on the $ x $ and $ y $ axes. The inner one with 30 uniform cells is between $ -2.5\times 10^8\cm $ and $ 2.5\times 10^8\cm $, the intermediate patch has 80 logarithmic cells up to $ |2.5\times 10^9|\cm $, and the last one includes 85 cells up to $ |R_\star| $. On the $ z $-axis we use one uniform patch of 2000 cells from $ z_{beg} $ to $ z = R_\star $. The total number of cells is then $ 2000 \times 360 \times 360 $. Our resolution along the jet axis is higher than that of \cite{lopez-camara2013} .}

\begin{figure}
	\includegraphics[scale=0.22]{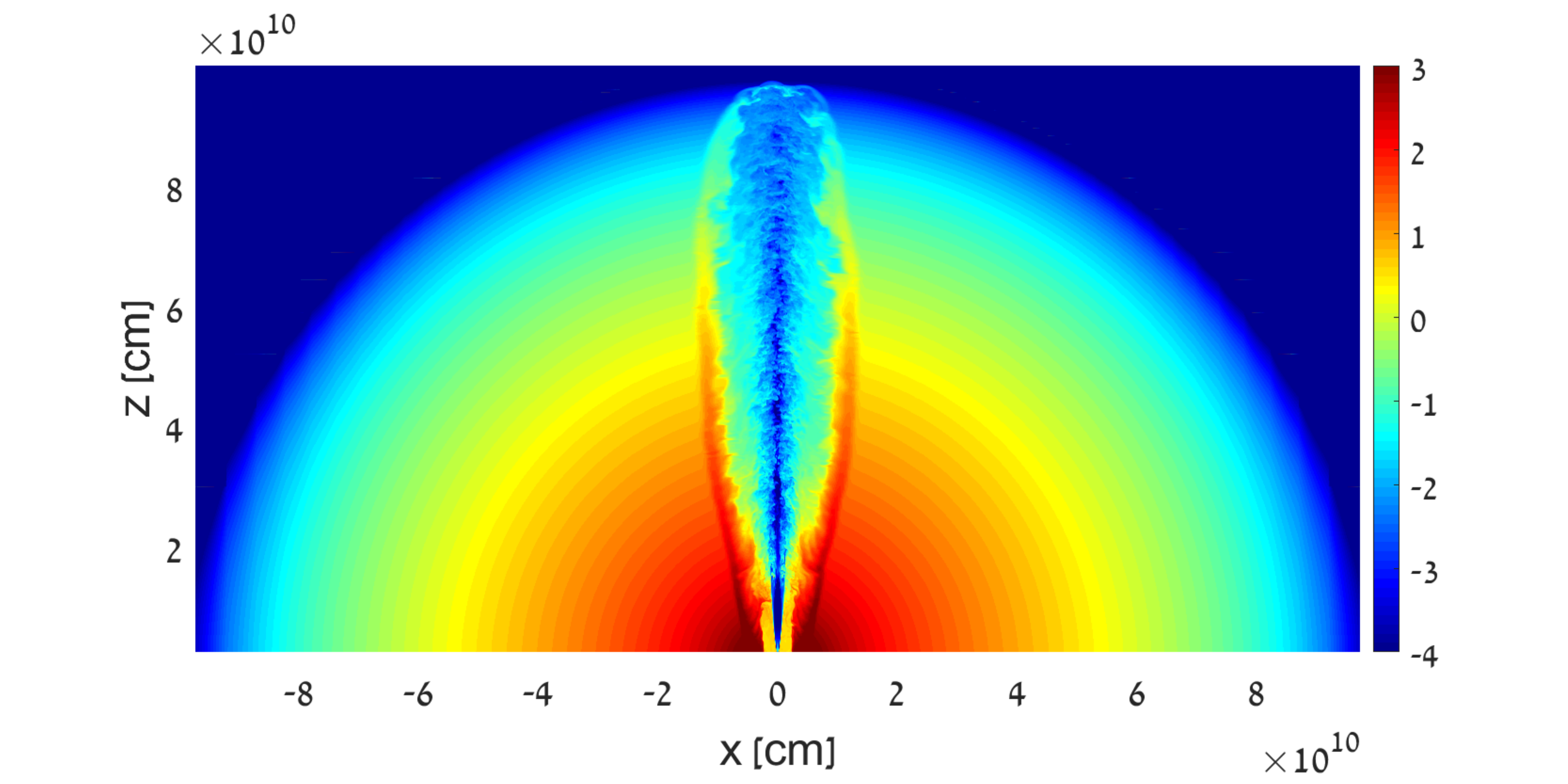}
	\caption{A logarithmic density $ (\rm{gr/cm^{3})} $ map of the jet inside the star at the breakout time.}    
	\label{fig:StarMap}
\end{figure}

\begin{figure}
\includegraphics[width=0.95\columnwidth]{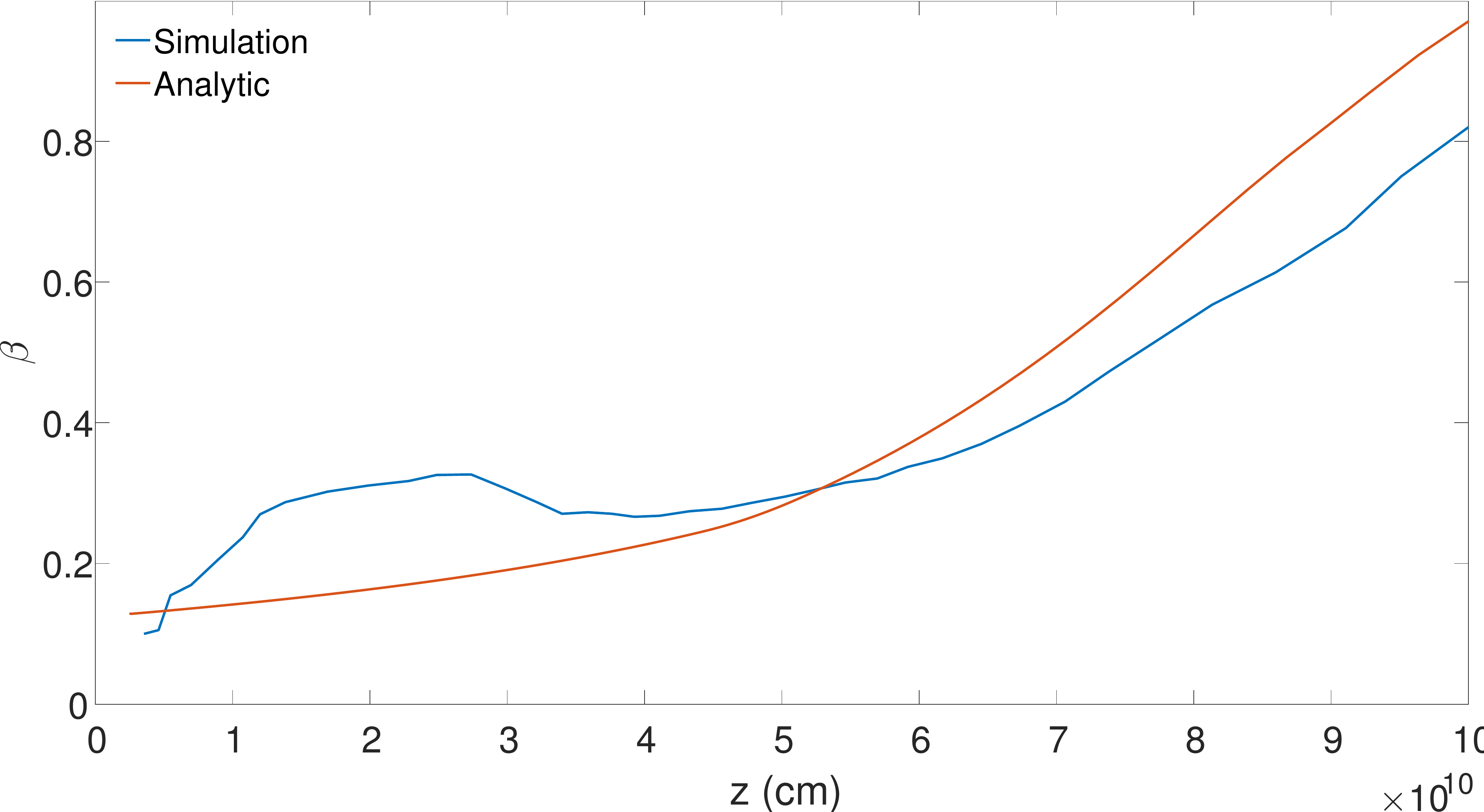}
\caption{The velocity as a function of the head location. The blue line presents the results of our 3D simulation and the red line is calculated semi-analytically using the calibrated numerical model. Note that once the head reaches about half way through the star the density drop is too sharp so the analytic model is not strictly applicable. Nevertheless, applying the integration of the semi-analytic model also to this region provide results that are not too different than the full numerical simulation. 
\label{fig:StarVZ}}
\end{figure}
	
Figure \ref{fig:StarMap} depicts a density map of the jet at breakout.
In Figure \ref{fig:StarVZ} we show a comparison between the simulation and semi-analytic model for the jet head location and velocity. The semi-analytic model is calculated according to scheme described in appendix \ref{app:SemiAnalyticModel}. The head velocity approaches Newtonian velocities ($\beta_s \approx 0.15c$) when it is deep within the star and it accelerates to relativistic velocities near the edge of the star where the stellar density drops sharply. At inner parts  ($z_h \lesssim 3 \times 10^{10}$) the numerical simulation is still affected by the initial conditions while in the outer parts of the star the density drop is too steep for the analytic model to be strictly applicable. Yet, the agreement is rather good.
{While the semi-analytic model predicts the breakout time to be $ 12 $s with the correction of $ N_s $, the simulation yields $ t_B = 10 $s.}
Using the latter for calibration we obtain that in the Newtonian regime the breakout time is:
\begin{equation}\label{eq:t_bo}
	t_B \approx 15\rm{s} ~ L_{j,iso,52}^{-1/3}\theta_{10^\circ}^{2/3}R_{11}^{2/3}M_{10}^{1/3}~= 17\rm{s} ~ L_{j,tot,50}^{-1/3}\theta_{10^\circ}^{4/3}R_{11}^{2/3}M_{10}^{1/3}~,
\end{equation}   
where $ L_{j,iso,52}$ is the isotropic equivalent jet luminosity (in units of $10^{52}$erg/s) and $L_{j,tot,50}$ is the total {\it bi-polar} luminosity of the jet (in units of $10^{50}$erg/s).  $ \theta_{10^\circ} = \theta_0/10^{\circ} $, $ R_{11}$ is the stellar radius in units of $10^{11}$cm and $ M_{10}$ is the progenitor mass in units of $ 10\msun $.
Note that for typical parameters the head velocity may be mildly relativistic where the accuracy of this formula is limited and the semi-analytic model we provide in the applet is more reliable. 

{Finally we compare the breakout time obtained by Eq. \ref{eq:t_bo} and by the semi-analytic model in the applet we provide to B11 and to the values found in previous numerical studies.  
The calibrated head velocity is a factor of $N_s^{5/(5-\alpha)}$ slower than the one obtained by B11 in the Newtonian regime (see Eq. \ref{eq:clibrated_bh}). Using $\alpha = 2$ to approximate the stellar profile, this implies a calibrated breakout time that is longer by a factor of $\approx 5$ compared to B11. This is also the difference between Eq. \ref{eq:t_bo} and equation 4 of \cite{bromberg11b}, which use B11 to calculate the  breakout time. When comparing our calibrated breakout time to previous numerical simulations we find an excellent agreement. \cite{mizuta2013} consider  a progenitor with a radius of $4 \times 10^{10}$cm and a mass of $13.95 M_\odot$. They inject a 2D jet with one sided luminosity $5 \times 10^{50}$ erg/s (i.e., $L_{j,tot,50}=10$), with various opening angles. They find breakout times of 4.5 s ang 7.5 s for opening angles of $8^o$ ($\Gamma_0=5$) and $16^o$ ($\Gamma_0=2.5$) respectively. For these parameters Eq. \ref{eq:t_bo} breakout times are 3.6 s and 9.1s while the semi-analytic model breakout times are 4.2 s and 8.3s.  \cite{lopez-camara2013} use the same progenitor as \cite{mizuta2013} and a 3D jet with one sided luminosity of $5.33 \times 10^{50}$ erg/s and an opening angle of $10^o$. They find in their highest resolution simulation a breakout time of 5.1 s compared to 4.8 s with Eq. \ref{eq:t_bo} and 5.1 s with the semi-analytic model.}

\section{Summary}
In this paper we use numerical simulations to test the validity, and find the accuracy, of the analytic model presented by B11 for the propagation of a relativistic unmagnetized jet in cold dense media. We do that by running a large set of 2D and 3D simulations to test the model in various settings and jet parameters.
We find that although being highly simplified, the model provides an excellent order of magnitude description for the structure and evolution of the jet-cocoon system. It provides a proper modeling for the various elements of the system (e.g., jet collimation, head velocity cocoon size and pressure, etc.), predicts their dependence on the jet and medium parameters as well as their temporal evolution accurately, and properly identifies the transition between various regimes (e.g., collimated to uncollimated). 

After finding that the power-law dependences of the model on the jet parameters and time are accurate we find the numerical coefficient of $\L$ (the dimensionless parameter that dictates the evolution) in 2D and 3D simulations and compare it to the analytic model. We find (as expected) that in the uncollimated regime the model agrees well with the simulations. However, in the collimated regime, especially when the head velocity is Newtonian, there are differences. We find that for the same system parameters and the same head location the value of $\sqrt{\L}$, which is approximately the head velocity in the Newtonian regime, is higher by a factor of about 3 in the simulations (both 2D and 3D) compared to the model of B11. When applying this correction factor we obtain an analytic model where the head velocity is accurate to within a factor of order unity. In the appendix we provide simple analytic formulae of the corrected model for the jet-cocoon evolution in regimes where these formulae are appropriate, i.e., fully Newtonian and fully relativistic head velocities in external medium with a power-law density profile. We also provide a link to an applet that calculates the jet-cocoon evolution semi-analytically for cases where the fully analytic formulae cannot be used. 

We compare some of the properties of 2D and 3D simulations. We find that in many aspects the 2D simulations capture the jet evolution properly, but there are some important differences. The general structure and evolution of 2D and 3D simulations are similar, as well the dependence on the jet and external medium parameters. The main morphological difference is in the structure of the head, where in 2D simulations there is a massive `plug' at the head front which divertes some of the jet elements sideways to dissipate their energy in oblique shocks. In contrast the head in 3D simulations is much less regular and there is no massive plug. When comparing the mixing of jet material in the cocoon we find that both 2D and 3D simulations show significant partial mixing at similar levels. However, the sites where the mixing takes place is different, suggesting that the mixing origin is different. Our findings suggest that 2D simulations can be used efficiently for the study of the general properties of the jet evolution, but cannot be trusted without 3D confirmation for the study of phenomena that depend on the head structure (e.g., the existence of a plug) or on mixing .

\section*{Acknowledgements}
We thank Omer Bromberg, Tsvi Piran and Re'em Sari for useful discussions. This research was partially supported by the I-Core center of excellence of the CHE-ISF, an ERC starting grant (GRB/SN), an ISF grant (1277/13) and an ISA grant.   	

\bibliographystyle{mnras}
\bibliography{Numericaljet}

\appendix
\section{}
\subsection{Integration parameters}\label{app:intParam}
We define the integration parameters used in the model,  $\zeta$, $\xi$, $\varrho$, and $\epsilon$, as follows: $z_h(t)=\int\beta_hdt'=\zeta t~ \beta_h(t)$, $r_c(t)=\int\beta_cdt'=\xi t~ \beta_c(t)$, $\bar{\rho}_a(z_h)=\int\rho_a~dV/V=\varrho\rho_a(z_h)$ and $\int\Gamma_h^{-2}dt'=\epsilon t~\Gamma_h(t)^{-2}$, which are defined collectively as,
\begin{equation}
 \Omega=\epsilon\varrho/\zeta\xi^2.
\end{equation} 
In the semi-analytic model the parameters are found by numerical integration\footnote{The value of $\varrho$ depends on the exact shape of the cocoon. In order to obtain a simple analytic expresion we approximate for that purpose that the cocoon is conical. Since we use the analytic expresion of $\Omega$ when calculating $\L_a$ (see Section \ref{sec:calibration}) the calibration factor $N_s$ corrects for this approximation as well.} . In the analytic model the value of the integration parameters depends on $\alpha$ and on weather the regime is Newtonian or relativistic. 
In the Newtonian regime ($\tilde{L}\ll 1$) we obtain $\xi=\zeta=(5-\alpha)/3$, $\varrho=3/(3-\alpha)$ and $\epsilon=1$. In the relativistic regime  ($\tilde{L} \gg 1$) we obtain $\xi=5/(3+\alpha)$, $\zeta=1$, $\varrho=3/(3-\alpha)$ and $\epsilon=5/(7-\alpha)$.

\subsection{Semi analytic model}\label{app:SemiAnalyticModel}
In the general case the model is semi-analytic, where the analytic jet equations should be propagated in time by a numerical integration. In \url{http://www.astro.tau.ac.il/~ore/propagation.html} we provide a java applet that carry out this semi-analytic integration. It provides the evolution with time of all the jet and cocoon parameters for any values of $L_j$, $\theta_0$ and $\rho_a(z)$ given by the user as well as several solutions for useful pre-defined jet configurations. Here we describe the algorithm of this applet.

To solve the set of equations semi-analytically, we propagate the jet parameters in small time steps. Given the jet parameters at time $t_i$ we propagate to time $t_{i+1}$ as follows. We use the head location at time $t_i$, $z_h(t_i)$, and the integration parameters up to that time, $\Omega (t_i)$, to calculate $\L_a$ using equations \ref{eq:Ltcol} or \ref{eq:Ltuncol} (depending on whether the jet is collimated or uncollimated). We find $\L_s$ using the calibration factor (see section \ref{sec:calibration}), $\L_s=N_s^2 \L_a$. Then we find the head velocity and cocoon expansion velocity by plugging $\L_s(t_i)$ and all other parameters at $t_i$ into equations \ref{eq:hvel}-\ref{eq:beta_c} and propagate the head location and cocoon size  as well as the integration parameters to $t_{i+1}$ using these velocities.

As an input the algorithm gets a density profile that starts at a minimal height $z_{base}$. To obtain the jet and cocoon properties at $z_{base}$ (i.e., the first time step), which are needed in order to start propagating them in time,  we assume that at $z<z_{base}$ the density profile is a power-law with $\alpha=2$ and find $\L_s$ in this region using equations \ref{eq:Ltcol} or \ref{eq:Ltuncol} and the calibration factor $N_s$. Having $\L_s$ we calculate all the other jet and cocoon properties at $z_{base}$, including $t_0=t(z_{base})$ and start evolving the algorithm from $t_0$.

\subsection{Analytic model}\label{app:analyticModel}
B11 provide analytic formulae in the various regimes for the case of a power-law density profile. Their formulae are written explicitly in terms of time, $t$, and external density at the location of the head, $\rho_a(z_h)$. Here we provide the same formulae in terms of $t$ only (the dependence on $\rho_a(z_h)$ is implicit), which is more useful for some applications. The dependence on $\rho_a(z_h)$ is eliminated by plugging the analytic expression of $z_h(t)$ into the density profile $\rho_a=A_\rho z^{-\alpha}$. In addition we add the calibration parameters in the Newtonian regime, where they significantly improve the model accuracy. \\

\noindent \textbullet {\it Collimated jet with a non-relativistic head \rm{[}$\tilde{L}<<1$\rm{]}}

In this regime we find that the analytic model needs calibration. The calibration factor varies between different quantities and depends on $\alpha$. Our best fit for the calibration factor is $N_s=0.35$.

\begin{equation}
z_h = N_s^{\frac{5}{5-\alpha}} \left(\frac{L_j}{A_\rho\theta_0^4}\right)^{\frac{1}{5-\alpha}} \left(\frac{2^4}{3^2\pi}\frac{(5-\alpha)^2}{(3-\alpha)}\right)^{\frac{1}{5-\alpha}}t^{\frac{3}{5-\alpha}}
\end{equation}


\begin{equation}\label{eq:clibrated_bh}
\beta_h = N_s^{\frac{5}{5-\alpha}} \frac{1}{c} \left(\frac{L_j}{A_\rho\theta_0^4}\right)^{\frac{1}{5-\alpha}} \left(\frac{2^43^{3-\alpha}}{\pi}\frac{(5-\alpha)^{\alpha-3}}{(3-\alpha)}\right)^{\frac{1}{5-\alpha}} t^{\frac{\alpha-2}{5-\alpha}}
\end{equation}


\begin{equation}
\begin{split}
P_c =& N_s^{\frac{3\alpha}{\alpha-5}} (L_j^{2-\alpha}A_\rho^{3}\theta_0^{2(1+\alpha)})^{\frac{1}{5-\alpha}}\\ &\left(\frac{3^3\pi^{\frac{\alpha}{2}-1}}{2^{1+\alpha}}\frac{(3-\alpha)^{\frac{\alpha}{2}-1}}{(5-\alpha)^{3}}\right)^{\frac{2}{5-\alpha}} t^{-\frac{4+\alpha}{5-\alpha}}
\end{split}
\end{equation}

\begin{equation}
\beta_c=\beta_h\theta_0\frac{1}{2\sqrt{\varrho}}
\end{equation}

\begin{equation}
r_c=z_h\theta_0\frac{1}{2\sqrt{\varrho}}
\end{equation}

\begin{equation}
\begin{split}
r_j=& N_s^{\frac{3\alpha}{2(5-\alpha)}} \left(\frac{L_j^3\theta_0^{4(2-\alpha)}}{A_\rho^{3}}\right)^{\frac{1}{2(5-\alpha)}}\\ &\left(\frac{2^{4(\alpha-2)}}{3^6\pi^3}(3-\alpha)^{2-\alpha}(5-\alpha)^6\right)^{\frac{1}{2(5-\alpha)}}\frac{1}{\sqrt{c}}t^{\frac{4+\alpha}{2(5-\alpha)}}
\end{split}
\end{equation}


\begin{equation}
\hat{z}=\frac{2r_j}{\theta_0}
\end{equation}

\vspace{0.3in}
\noindent \textbullet {\it Collimated jet with a relativistic head \rm{[}$1 \ll \tilde{L}<\left(\frac{16\Omega}{3\theta_0^2}\right)^{2/3}$\rm{]}}


\begin{equation}
z_h\sim ct
\end{equation}

\begin{equation}
\beta_h\sim 1
\end{equation}

\begin{equation}
\Gamma_h=\left(\frac{L_j}{A_\rho c^{5-\alpha}\theta_0^4}\right)^{1/10}\left(\frac{1}{10\pi}\frac{(3+\alpha)^2}{(3-\alpha)(7-\alpha)}\right)^{1/10}t^{\frac{\alpha-2}{10}}
\end{equation}


\begin{equation}
P_c=\left(\frac{L_j^2 A_\rho^3 \theta_0^2}{c^{3\alpha}}\right)^{1/5}\left(\frac{1}{10\pi}\frac{(3+\alpha)^2}{(3-\alpha)(7-\alpha)}\right)^{2/5}t^{-\frac{3\alpha+4}{5}}
\end{equation}

\begin{equation}
r_c=\left(\frac{L_j\theta_0c^{\alpha}}{A_\rho}\right)^{1/5}\left(\frac{5^4}{3^4 2\pi}\frac{(3-\alpha)^3}{(7-\alpha)(3+\alpha)^3}\right)^{1/5}t^{\frac{3+\alpha}{5}}
\end{equation}

\begin{equation}
\beta_c=\left(\frac{L_j\theta_0}{A_\rho c^{5-\alpha}}\right)^{1/5}\left(\frac{1}{3^410\pi}\frac{(3-\alpha)^3(3+\alpha)^2}{(7-\alpha)}\right)^{1/5}t^{\frac{\alpha-2}{5}}
\end{equation}

\begin{equation}
r_j=\left(\frac{L_j^3\theta_0^8}{A_\rho^3c^{5-3\alpha}}\right)^{1/10}\left(\frac{5}{2^4\pi^{3/2}}\frac{(3-\alpha)(7-\alpha)}{(3+\alpha)^2}\right)^{1/5}t^{\frac{3\alpha+4}{10}}
\end{equation}



\begin{equation}
\hat{z}=\frac{2r_j}{\theta_0}
\end{equation}

\vspace{0.3in}
\noindent \textbullet {\it Uncollimated jet with a causally connected relativistic head \rm{[}$\left(\frac{16\Omega}{3\theta_0^2}\right)^{-2/3} \ll \tilde{L}<\theta_0^{-4}$\rm{]}}\\
The jet has a conical structure such that the collimation shock remains open and $\theta_j \approx \theta_0$.

\begin{equation}
z_h\sim ct
\end{equation}

\begin{equation}
\beta_h\sim 1
\end{equation}

\begin{equation}
\Gamma_h=\left(\frac{L_j\theta_0}{4\pi A_\rho c^{5-\alpha}}\right)^{1/4}t^{\frac{\alpha-2}{4}}
\end{equation}


\begin{equation}
P_c=\left(\frac{L_j\theta_0^2A_\rho^3}{c^{3(\alpha-1)}}\right)^{1/4}\left(\frac{1}{5\pi}\frac{(3+\alpha)^2}{(3-\alpha)(7-\alpha)}\right)^{1/2}t^{-\frac{3\alpha+2}{4}}
\end{equation}

\begin{equation}
\beta_c=\left(\frac{L_j\theta_0^2}{A_\rho c^{5-\alpha}}\right)^{1/8}\left(\frac{1}{3^25\pi}\frac{(3-\alpha)(3+\alpha)^2}{(7-\alpha)}\right)^{1/2}t^{\frac{\alpha-2}{8}}
\end{equation}

\begin{equation}
r_c=\left(\frac{L_j\theta_0^2c^{\alpha+3}}{A_\rho}\right)^{1/8}\left(\frac{5^3}{3^2\pi}\frac{(3-\alpha)}{(7-\alpha)(3+\alpha)^2}\right)^{1/2}t^{\frac{\alpha+6}{8}}
\end{equation}

\begin{equation}
\theta_c=\beta_c
\end{equation}

\subsection{Numerical convergence}\label{app:convergence}
We test for convergence in the jet evolution for 2D and 3D simulations by varying the jet resolution
for the initial conditions of simulations 12 and 3Df. Both simulation has $\alpha=2$ and therefore the jet head velocity in every simulation is expected to reach its converged value after enough time. Since all physical scales are growing linearly with time for $\alpha=2$, we expect that the time for convergence will increase linearly with the mesh size. Below we first discuss convergence in 2D, and then in 3D. 

\begin{figure}
\begin{center}
\includegraphics[width=1\columnwidth]{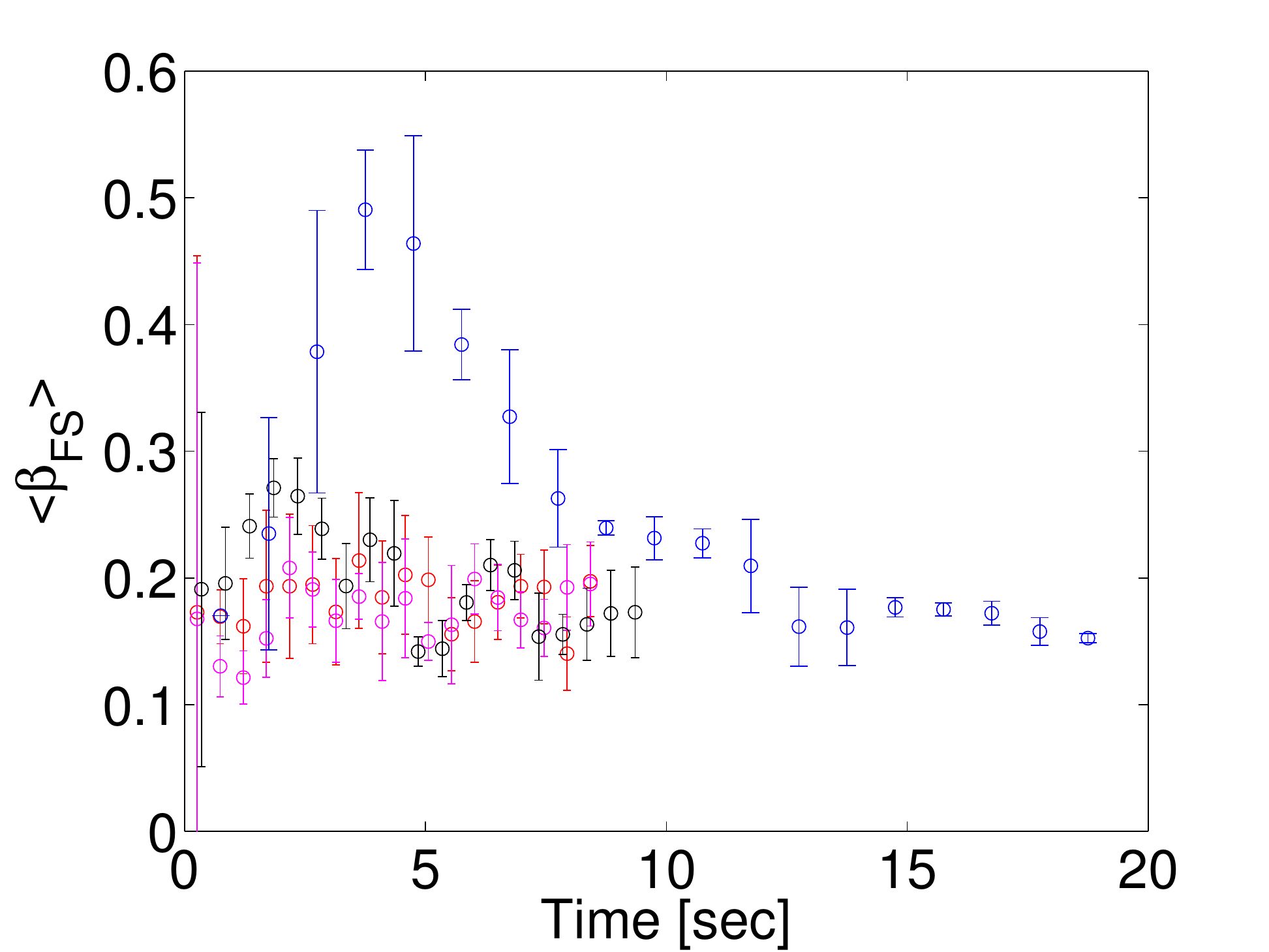}
\caption{Results of the 2D convergence test showing the head forward shock velocity as a function of time for simulation 12 in four different resolutions. The resolutions are: A) 200 x 2016 (blue), B) 400 x 4032 (black), C) 800 x 8064 (red), and D) 1600 x 16128 (magenta).
\label{fig:converge}}
\end{center}
\end{figure}

In 2D we consider mesh structure that has uniform distribution for $0~<~r~<2\times10^{9}\mathrm{cm}$ and $1\times10^9\mathrm{cm}~<~z~<4\times10^{10}\mathrm{cm}$ with logarithmic distribution for $2\times10^{9}\mathrm{cm}~<~r~<1\times10^{11}\mathrm{cm}$ and $4\times10^{10}\mathrm{cm}~<~z~<4\times10^{11}\mathrm{cm}$. We consider four levels of meshes refinement in $r \times z$: A) 200 x 2016, B) 400 x 4032, C) 800 x 8064, and D) 1600 x 16128. We run each simulation for $10\mathrm{s}$ ($20\mathrm{s}$ for case A) and use the head velocity to investigate convergence. Figure \ref{fig:converge} presents the head velocity as a function of time in the four simulations. Lower resolution simulations have at first a faster head (similarly to the findings of \citealt{mizuta2013}), but after enough time all simulations converge to the same head velocity. As expected the time of convergence is roughly linear with the mesh size where in simulation D the head velocity converges within the first second, in C after $\sim 1-2\mathrm{s}$, in B after $\sim 5 \mathrm{s}$ and in  A after $\sim12\mathrm{s}$. For our study we use resolution B in all 2D simulations. We run each simulation for 8s first and check if its head velocity stabilised. If not we continue to run until it does. In all runs we also requie that the height of the jet head is large enough ($z_h> 1.2 \times 10^{10}\mathrm{cm}$) and that the jet is wide enough so the values of $z_{base}$ and $r_{noz}$ will not affect the results (see below).

\begin{figure}
\begin{center}
\includegraphics[width=1\columnwidth]{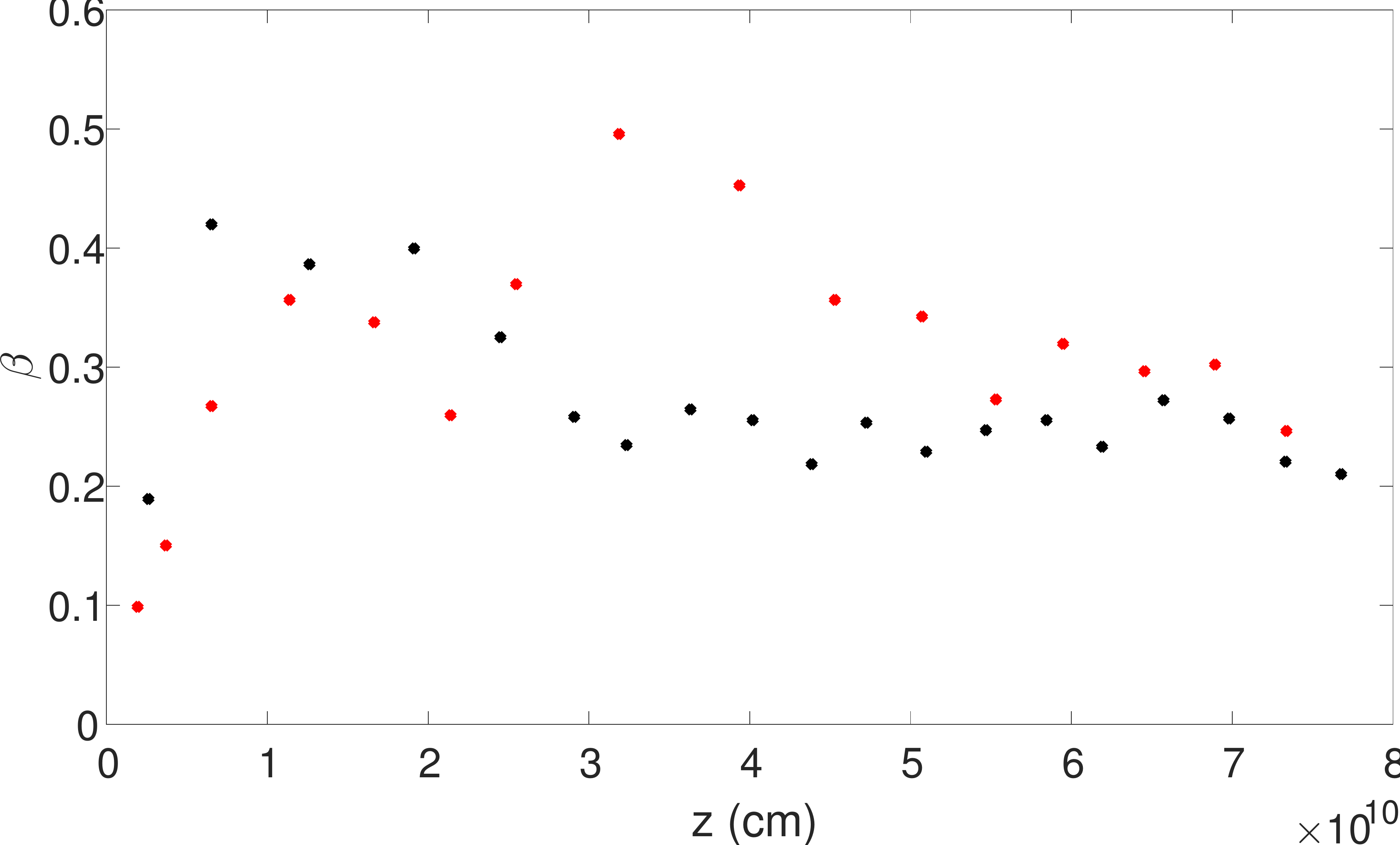}
\caption{Convergence test for simulation 3Df depicting the head velocity  as a function of the head location. The original cells size is presented in black, while in red the cells size is doubled.
Each dot presents an average over 0.5s. As expected, both simulations are converged to the same velocity after some time, where the head location upon convergence of the lower resolution simulation is approximately at twice the height compared its location when the high resolution simulation converges.
\label{fig:conv3d}}
\end{center}
\end{figure}

For the numerical convergence in 3D we make use of the highest resolution model 3Df described in  section \ref{sec:num}  and a test simulation with half the resolution, i.e., the cells are twice as big with a total number of grid cells of $ 250 \times 250 \times 770 $. Figure \ref{fig:conv3d} depicts the 
evolution of the head velocity as a function of head location. Similarly to the 2D simulations, both resolutions show an episode of evolving velocity at early time  that is converged to the same constant velocity at later time. As expected, the time at which the lower resolution is converged is roughly twice that of the high resolution convergence, so the time of convergence is approximately linear at the cells size.

\subsection{The effect of nozzle size and base height}\label{app:nozzle}
Across all simulations the nozzle size must satisfy $r_{noz}<r_{col}$, where $r_{col}$ being the radius of the collimation shock. We consider different nozzle sizes and find that as long as the resolution allows roughly 10 meshes across the nozzle, and $r_{noz}<<r_{col}$, then the simulation is independent of the nozzle size. 
The base height may affect the jet evolution if the collimation shock is not high enough so the jet cannot fully expand to the point where its opening angle remains constant before collimation starts. We find, as expected, that as long as the jet collimation shock in the collimated regime or jet head in the uncollimated regime, are far above the nozzle (i.e., $z_h,\hat{z}>>z_{base}$) the jet evolution does not depend on the base height. 

\subsection{The effect of the equation of state}\label{app:eos}
\begin{figure}
\begin{center}
\includegraphics[width=1\columnwidth]{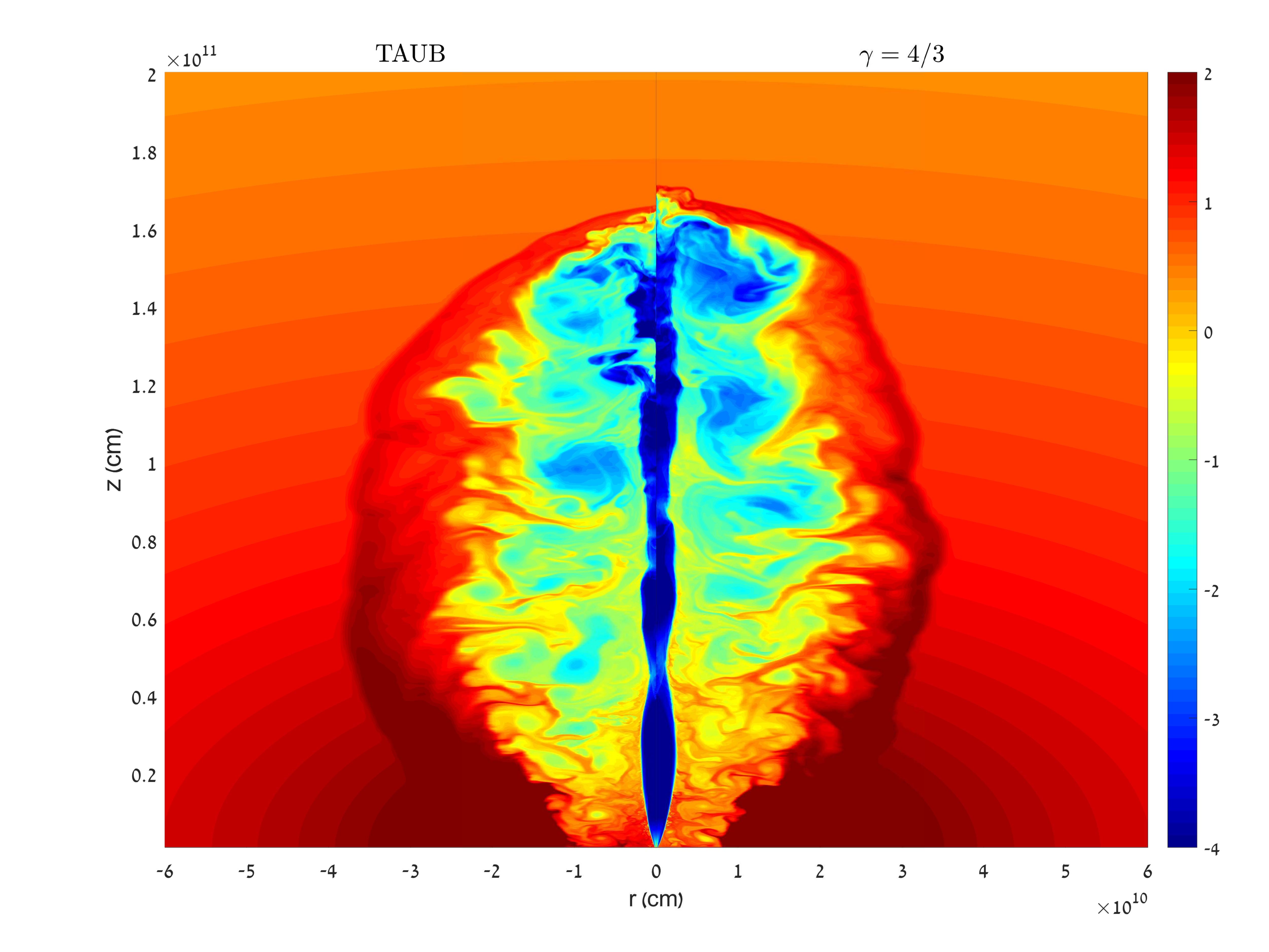}
\caption{A density map comparing between two simulations of model 12 with different equations of state, which is a polytopic index of 4/3 on the right and Taub on the left. The two snapshots, which where taken at the same time, are very similar. For example, the difference in the head velocity is less than 5\%.}
\label{fig:EOS}
\end{center}
\end{figure}

Throughout this work our simulations use an ideal equation of state with a constant polytropic index of $4/3$, as appropriate for a radiation dominated gas. To verify that our results are not strongly affected in case that the gas pressure is not dominated by radiation we carry out also a simulation with a Taub equation of state \citep{mignone2005}, which vary smoothly between the Newtonian (polytropic index of 5/3) and relativistic (polytropic index of 4/3) equations of state, according to the temperature, assuming that the pressure is dominated by the gas. We tested a collimated jet with a Newtonian head where some of the shocked gas is not relativistically hot. As can be seen in figure \ref{fig:EOS}, we find no significant difference in the simulation properties between the different equations of state.

\end{document}